# Structural Solutions for Cross-Layer Optimization of Wireless Multimedia Transmission

Fangwen Fu and Mihaela van der Schaar

Department of Electrical Engineering, University of California, Los Angeles, Los Angeles, CA, 90095

{fwfu, mihaela}@ee.ucla.edu

## ABSTRACT

In this paper, we propose a systematic solution to the problem of cross-layer optimization for delay-sensitive media transmission over time-varying wireless channels as well as investigate the structures and properties of this solution, such that it can be easily implemented in various multimedia systems and applications. Specifically, we formulate this problem as a finite-horizon Markov decision process (MDP) by explicitly considering the users' heterogeneous multimedia traffic characteristics (e.g. delay deadlines, distortion impacts and dependencies etc.), time-varying network conditions as well as, importantly, their ability to adapt their cross-layer transmission strategies in response to these dynamics. Based on the heterogeneous characteristics of the media packets, we are able to express the transmission priorities between packets as a new type of directed acyclic graph (DAG). This DAG provides the necessary structure for determining the optimal cross-layer actions in each time slot: the root packet in the DAG will always be selected for transmission since it has the highest positive marginal utility; and the complexity of the proposed cross-layer solution is demonstrated to linearly increase w.r.t. the number of disconnected packet pairs in the DAG and exponentially increase w.r.t. the number of packets on which the current packets depend on. The simulation results demonstrate that the proposed solution significantly outperforms existing state-of-the-art cross-layer solutions. Moreover, we show that our solution provides the upper bound performance for the cross-layer optimization solutions with delayed feedback such as the well-known RaDiO framework.

**Keywords**: Multimedia Streaming, Wireless Multimedia Networking, Delay Sensitive Communication, Markov Decision Process, Directed Acyclic Graph.

## I. INTRODUCTION

Existing wireless networks provide dynamically varying resources with only limited support for the Quality of Service (QoS) required by the *delay-sensitive, bandwidth-intense and loss-tolerant* multimedia applications. One of the key challenges associated with the robust and efficient multimedia transmission over wireless networks is the *dynamic* characteristics of both the wireless networks and multimedia sources experienced by the wireless user [1]. To overcome this challenge, cross-layer optimization has been extensively investigated in recent years in order to maximize the application's utility (e.g. multimedia applications) given the underlying time-varying and error-prone network characteristics. A brief summary of this research is provided next.

Existing cross-layer optimization solutions often involve only the layers below the application layer, which collectively aim to maximize QoS metrics such as throughput, packet loss rate, average or worst case delay etc., but without considering the specific characteristics and requirements of the applications. For example, in [2][5], a method is proposed for minimizing the incurred average delay under energy (or average power) constraints for applications where packets have the same distortion impact. In [4], the optimal packet scheduling algorithm is developed for the transmission of a group of equal-importance packets, which minimizes the consumed energy while satisfying their common delay deadline. This packet scheduling algorithm is further extended in [3] to the case in which each packet has its own delay constraints. However, the above papers disregard key properties of multimedia applications: the interdependencies among packets and their different distortion impacts.

To take into consideration the heterogeneous characteristics of the multimedia data, one solution is to employ Unequal-Error-Protection (UEP) techniques [10][11] using Forward Error Control (FEC) to differentially protect the video packets based on their distortion impacts, delay deadlines and packets' dependencies. However, these solutions assume only simplistic underlying network (channel) models (constant transmission rate and packet loss rate) and they do not consider the time varying channel conditions and the adaptation of transmission parameters at the other layers of the network stack, besides the application (APP) layer.

By considering the time-varying channel conditions, packet scheduling is employed to schedule the packet transmission based on both the heterogeneous characteristics of media data and the time-varying channel conditions. In [8][9], the packets are scheduled for transmission over a constant channel (with constant packet error rate) in order to minimize the application distortion while satisfying the delay constraint. However, these solutions do not take into account the complicated dependencies between media packets and the adaptation capabilities at the Media Access Control (MAC) and Physical (PHY) layers. By enabling the adaptation capabilities at the MAC and PHY layers, the work in [6] developed a cross-layer optimization solution which is able to schedule the video packets over the time-varying wireless channel according to their various distortion impacts. However, this cross-layer optimization only considers the wireless channel conditions observed at the current time, without considering or modelling future transmission opportunities which may exhibit different channel conditions. In [7], the complicated dependencies between the multimedia packets are expressed as a DAG and the packet scheduling is optimized under a rate-distortion framework (named RaDiO), which takes into consideration the heterogeneous characteristics of multimedia data. However, RaDiO disregards the dynamics and error protection capabilities available at the lower layers of the protocol stack (e.g. MAC and PHY layers). In summary, a systematic cross-layer optimization framework for media communication over time-varying wireless networks is still missing.

To overcome this challenge, in this paper we develop a cross-layer optimization framework for single-user multimedia transmission over single-hop wireless networks by explicitly considering the heterogeneous characteristics of multimedia data, time-varying network conditions and adaptation capability of the user at the various layers of the protocol stack. We jointly optimize[1] the packet scheduling at the APP layer and transmission strategy (e.g. retransmission, power allocation and modulation selection) adaptation at the MAC and PHY layers.

Specifically, we first consider the cross-layer optimization for a single-packet transmission and formulate it as a finite-horizon optimal stopping problem in which the threshold-based cross-layer transmission policy is determined. The threshold is computed based on the delay deadline, distortion impact and underlying time-varying network conditions, and represents the future net utility determined by evaluating future potential transmission opportunities. The threshold is decreased when a packet approaches the delay deadline and the marginal utility of this packet (i.e. current utility minus the threshold) correspondingly increases. Hence, when the packet is closer to its delay deadline, it will have a higher chance to be transmitted.

We then extend the cross-layer optimization for the single packet to multiple packets, each having different attributes (e.g. arrival times, delay deadlines, distortion impact and dependencies). In addition to exploiting future potential transmission opportunities (e.g. by setting a decreasing threshold) for each packet, we also have to consider the mutual impact among multiple packets (i.e. determining which packets should be transmitted first) due to their dependencies and their sharing of the same transmission resource (e.g. transmission power etc.). To do this, we define the transmission priorities between the packets based on their attributes, and express the transmission priorities as a DAG, which can be viewed as an augmented DAG expression of the packet dependencies proposed in [7]. The proposed DAG expression of the packets' priorities provides the necessary structure for determining the optimal cross-layer actions at each time slot. Specifically, we will always select the root packet in the DAG to transmit since it has the highest marginal utility. We show that the complexity of the cross-layer optimization using the DAG linearly increases with the number of disconnected packet pairs (i.e. packets which cannot be prioritized) and exponentially increases with the number of dependency states (i.e. packets on which the packets to be transmitted depend), both of which are determined by the media characteristics.

Although the structural solution to the cross-layer optimization is developed by knowing the packet transmission outcomes before scheduling the next packet[2], the proposed framework can be extended to the case in which the transmission outcomes are delayed as considered in RaDiO [7]. The extension can be performed by reformulating the cross-layer optimization as a partially observable MDP (POMDP) [23] in which the probabilistic traffic state (representing whether the media packets are transmitted or not) can be updated based on the feedback at each time slot. However, we leave this interesting problem for future investigation and focus instead on the cross-layer optimization with immediate transmission outcome. However, we know that the imperfect observation at each time slot on the packet transmission outcome leads to degradations in the received media quality, and hence, our cross-layer solution gives an upper bound on the performance of current state-of-art cross-layer solutions that use delayed acknowledgements. Moreover, different from RaDiO, where only the mutual impact among the packet dependencies are considered (i.e. a linear transmission cost is assumed), our cross-layer solution provides a systematic framework to characterize the mutual impact among the media packets based on their packet dependencies as well as their sharing of the same transmission resource. In summary, the differences between RaDiO and our proposed solution are listed in Table 1.

---

[1] We use User Datagram Protocol (UDP) in the transport layer.

2 In this single-hop wireless network, the transmission of each packet can be acknowledged in the MAC layer, e.g. in IEEE 802.11 [20]. Hence, the transmission outcome (i.e. successful transmission or lost packet) of each packet is known to the transmitter before the transmission of the next packet.

Table 1. Comparison between RaDiO and proposed cross-layer optimization solution

| | **RaDiO [7]** | **Proposed solution** |
|---|---|---|
| **Consideration of heterogeneous multimedia data** | Yes | Yes |
| **Feedback of transmission outcome** | Delayed acknowledgement | Immediate acknowledgement |
| **Transmission cost form** | Linear | convex |
| **Adaptation** | Single layer adaptation | Cross-layer adaptation |
| **Performance (under time-varying wireless channel)** | Suboptimal | Optimal |

The paper is organized as follows. Section II characterizes the attributes of the multimedia traffic. Section III formulates the single-packet cross-layer optimization as an optimal stopping problem and proposes a novel threshold-based scheduling policy. Section IV formulates the transmission of multiple independently decodable packets as an MDP and presents structural properties of the corresponding solutions. Section V formulates the transmission of multiple interdependent packets as an MDP by considering the dependencies between them and characterizes corresponding structural solutions. Section VI presents the simulation results, which is followed by the conclusions in Section VII.

## II. A BRIEF OVERVIEW OF MULTIMEDIA TRAFFIC CHARACTERISTICS

In this section, we discuss how the heterogeneous attributes of multimedia traffic[3] can be modelled. In the past work, multimedia traffic (e.g. video traffic) is often modelled as a leaky bucket with constraints (e.g. peak rate constraint, delay constraint etc.) [17]. However, these models only characterize the rate change in multimedia traffic and its corresponding impact on the average delay. They do not explicitly consider the heterogeneous characteristics of the multimedia traffic. As in [14][15][22], the multimedia data are often encoded interdependently, using sophisticated prediction-based coding solutions, in order to remove the temporal correlation existing among the data. This introduces sophisticated dependencies between the encoded data across time. In [7], multimedia traffic is modelled using a DAG, which takes into account the distortion impact and delay deadline of each packet, as well as the inter-dependencies among packets, thereby accurately capturing the time-varying traffic characteristics. In this paper, we also use a DAG to characterize the traffic. The encoded data is packetized into multiple data units (DUs). For example, for video applications, the DUs are video packets, video frames etc. The DU's attributes are listed below:

- *Size:* The size of DU $j \in \mathbb{N}$ is denoted as $l_j$ (measured in bits).
- *Distortion impact*: Each DU $j$ has a distortion impact $q_j$, which is the amount by which the video distortion will be reduced if the DU is decoded at the destination.
- *Arrival time:* The arrival time is the time at which the DU is ready for transmission. The arrival time for DU $j$ is denoted by $t_j$. If the video data is pre-encoded, then each DU is available for transmission at $t_j = 0$. If the video data is encoded in real time, the arrival time is the time when the DU is packetized and injected into the post-encoding buffer.
- *Delay deadline*: The delay deadline is the time by which the data unit must be decoded and displayed. If the DU is not received at the destination by the delay deadline, it will be discarded and it will be considered useless[4]. The delay deadline is denoted by $d_j$ and $t_j < d_j$, since the DU needs to be transmitted before its expiration.
- *Dependency*: The dependencies among the DUs are expressed as a DAG as in [7]. In this paper, we assume that, if DU $j'$ depends on DU $j$ (i.e. there exists a path directed from DU $j'$ to DU $j$ and denoted by $j \prec j'$), then $t_j \le t_{j'}$ and $d_j \le d_{j'}$. In other words, DU $j$ should be encoded and decoded prior to DU $j'$. If DU $j$ is not successfully transmitted prior to the delay deadline, then all the DUs depending on DU $j$ will be considered useless.

During the transmission, each DU is packetized into one (or multiple) packet(s). Our cross-layer optimization is performed at the packet level. With abuse of notation, we consider that each packet $j$ (instead of DU $j$) has size $l$, distortion impact $q_j$, arrival time $t_j$, delay deadline $d_j$ and its dependencies to other packets are expressed by a DAG[5]. In the following, we will first consider the cross-layer optimization for a single packet in Section III, and then describe the cross-layer optimization for multiple packets which can be independently decoded in Section IV. Finally, we will present the cross-layer optimization for multiple packets which are interdependent in Section V.

---

[3] Video traffic can be generated in real time or be pre-encoded.

## III. MDP FOR SINGLE PACKET TRANSMISSION

In this section, we propose an MDP formulation and corresponding solutions for a single packet transmission. Without loss of generality, let us consider a packet $j$ with $t_j = 0$. The packet can be scheduled for transmission at time slots $0,1,\cdots,d_j$. We define a scheduling policy $\pi_j = \left(\pi_j^0,\cdots,\pi_j^{d_j}\right) \in \{0,1\}^{d_j+1}$, whose components $\pi_j^t$ represent the scheduling action taken in time slot $t$: $\pi_j^t = 1$, if the packet is scheduled to be transmitted, and $\pi_j^t = 0$, otherwise. The packet scheduling policy is performed at the APP layer.

In each time slot $t$, the user experiences a channel condition $h^t \in \mathcal{H}$, where $\mathcal{H}$ is the set of finite possible network conditions (or channel conditions). In this paper, we assume that the channel condition $h^t$ can be modelled as finite state Markov chain (FSMC) [16] with transition probability $p_h\left(h' \mid h\right) \in [0,1]$. At time slot $t$, if the packet is scheduled to be transmitted (i.e. $\pi_j^t = 1$), the wireless user deploys the transmission strategy $a_j^t \in \mathcal{A}$, where $\mathcal{A}$ is the set of possible transmission strategies available for the user. The transmission strategies can include the modulation and channel coding selection or power allocation at the physical layer, or retransmission at the MAC layer. The incurred transmission cost (e.g. the amount of transmission time or the amount of power allocated) is denoted by $\rho^t\left(l,h^t,a_j^t\right)$.

The objective of the cross-layer optimization for the single packet transmission is to maximize the discounted net utility, i.e.

---

[4] In real multimedia applications, the discard data can be concealed using previous received data. The error concealment algorithm can be easily incorporated into our proposed cross-layer optimization framework. In this paper, we do not consider such concealment algorithms at the decoder side.

[5] If the DU is divided into multiple packets, then the dependencies between these packets can be expressed as a dependency chain.

$$\max_{\pi_j^t \in \{0,1\}, a_j^t \in \mathcal{A}} \sum_{t=0}^{d_j} \alpha^t \left( q_j - \lambda \rho^t \left( l, h^t, a_j^t \right) \right) \pi_j^t$$

$$s.t. \sum_{t=0}^{d_j} \pi_j^t \le 1 \qquad (1)$$

$$r\left(h^t, a_j^t\right) = \pi_j^t l,$$

where $r\left(h^t, a_j^t\right)$ is the average amount (in bits) of successfully transmitted data at time slot $t$, which is an increasing function of $a_j^t$. $\alpha \in [0,1]$ is the discount factor. The reason why we introduce the discount factor here will be made clear in Section B. In this paper, we assume that the wireless user is able to transmit multiple packets within one time slot, which will be detailed in Section IV. The first constraint in Eq. (1) means that the packet is scheduled for transmission within one time slot. This is because we assume that the packet scheduling is performed at the APP layer and we do not need to consider APP layer retransmissions, which can be instead implemented more efficiently at the MAC layer in this one-hop wireless network. The second constraint means that, once the packet is scheduled for transmission at time slot $t$, i.e. $\pi_j^t = 1$, the transmission strategy $a_j^t$ is selected such that the total number of successfully transmitted bits equals the length of the packet. Note that the packet scheduling $\pi_j^t$ and transmission strategy $a_j^t$ need to be jointly optimized in order to maximize the net utility.

In the following, we provide two examples of cross-layer optimization and show how to compute $r\left(h^t, a_j^t\right)$ and $\rho^t\left(l, h^t, a_j^t\right)$.

***Example 1*** (Application-MAC-PHY cross-layer optimization): A wireless user is scheduling packets for transmission over a single-hop wireless network by optimizing the packet scheduling policy and the number of retransmissions. In each time slot, the wireless user experiences a channel condition $h^t$, which represents the Signal-to-Noise Ratio (SNR)[6] of the wireless channel. The transmission rate and packet loss probability are given by $R\left(h^t\right)$ and $p_L\left(h^t\right)$, which can be computed as in [19] given a selected modulation and channel coding schemes. At the MAC layer, a TDMA-like channel access protocol is assumed and the wireless user retransmits packets if they fail to be received by the destination. The transmission strategy $a_j^t \in \mathbb{R}_+$ is defined as the average number of transmissions until the packet is received successfully. Once the packet is scheduled for transmission at time slot $t$, i.e. $\pi_j^t = 1$, the transmission strategy is selected as $a_j^t = 1/(1 - p_L(h^t))$. The transmission cost is the average amount of time spent on the transmission of packet $j$ and is given by $\rho^t(l, h^t, a_j^t) = a_j^t l / R(h^t)$. It is clear that the transmission cost is a linear function of the number of transmitted bits.

[6] In this example, we assume that the wireless user transmits the data using constant transmission power.

***Example 2*** (Application-PHY cross-layer optimization) A wireless user is scheduling packets for transmission over a single-hop wireless network by optimizing the packet scheduling policy and power allocation. In each time slot, the wireless user experiences a channel condition $h^t$ which represents the channel gain of the wireless channel. The transmission rate is given by $r(h^t, a_j^t) = \frac{1}{2}\log(1 + h^t a_j^t)\Delta T$, where $a_j^t$ represents the amount of power allocated for transmission and $\Delta T$ is the length of the time slot. Once the packet is scheduled at time slot $t$, i.e. $\pi_j^t = 1$, the transmission strategy is selected as $a_j^t = \left(2^{2l/\Delta T} - 1\right)/h^t$. The transmission cost is given by $\rho^t(l, h^t, a_j^t) = a_j^t = \left(2^{2l/\Delta T} - 1\right)/h^t$. It can be shown that the transmission cost function is a convex function of the number of transmitted bits [18].

### A. Optimal Stopping Problem Formulation for Single-packet Transmission

In this section, we will show how to formulate the cross-layer optimization defined in Eq. (1) as a finite horizon optimal stopping problem [13], which is a special MDP. Specifically, we define the traffic state of the packet at time slot $t$ as $b_j^t \in \{0,1\}$, where $b_j^t = 0$ if the packet is successfully received by the destination, and otherwise $b_j^t = 1$. The transition of the traffic state $b_j^t$ is illustrated in Figure 1, where the black dot represents $b_j^t = 0$ (i.e. terminating state) and the circle represents $b_j^t = 1$. When the state is $b_j^t = 0$, then the transmission of the packet is stopped. When the current state is $b_j^t = 1$, it will transit to $b_j^{t+1} = 0$ in the next time slot and incur transmission cost $\rho_j^t$, if the scheduling action is $\pi_j^t = 1$ [7]; otherwise, it will transit to $b_j^{t+1} = 1$ and have no transmission cost $\rho_j^t = 0$. When the current state is $b_j^t = 0$, it will always transit to $b_j^{t+1} = 0$ without any transmission cost, i.e. $\rho_j^t = 0$. It is clear that the transition of the traffic state $b_j^t$ is Markovian. We note that the trellis in Figure 1 is a simplified version of the trellis developed in [7] when having immediate feedback for the transmission outcomes.

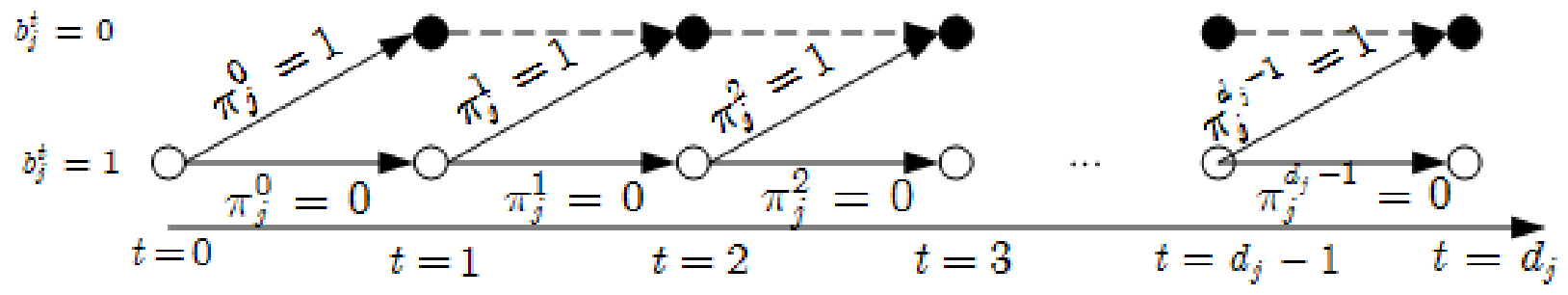

Figure 1. Transition of the traffic state $b_j^t$ under the scheduling policy

We define the state of packet as $s_j^t = \left(b_j^t, h^t\right)$, which include the traffic state $b_j^t$ and channel state $h^t$. Since both $b_j^t$ and $h^t$ are Markovian, the transition of the state $s_j^t$ is also Markovian and the state transition probability is given[8] by $p\left(s_j^{t+1} \mid s_j^t, \pi_j^t\right) = p\left(b_j^{t+1} \mid b_j^t, \pi_j^t\right) p\left(h^{t+1} \mid h^t\right)$.

**Lemma 1**. The cross-layer optimization for the single-packet transmission is a $\left(d_j + 1\right)$-stage[9] optimal stopping problem with the state $s_j^t$, state transition probability $p\left(s_j^{t+1} \mid s_j^t, \pi_j^t\right)$, action $\left(\pi_j^t, a_j^t\right)$, and immediate net utility

$$u_j^t\left(s_j^t, \left(\pi_j^t, a_j^t\right)\right) = \begin{cases} 0 & if\ t = d_j + 1\ or\ b_j^t = 0 \\ \left(q_j - \lambda \rho^t\right)\pi_j^t & if\ t \neq d_j + 1\ or\ b_j^t \neq 0 \end{cases}.$$

The proof is straightforward based on the discussion above, and thus it is omitted here due to space limitations. In this optimal stopping problem formulation, the traffic state $b_j^t = 0$ is the terminating state in which the transmission will be stopped since the packet is successfully received. It is clear that, by formulating the cross-layer optimization problem in Eq. (1) as an optimal stopping problem, the constraint on the packet scheduling is automatically satisfied. This is because, once the packet is scheduled for transmission, the packet will be in the stopping state and will not be transmitted in the future. Once the packet is scheduled for transmission at time slot $t$, the optimal transmission strategy $a_j^{t,*}$ is selected such that $r\left(h^t, a_j^{t,*}\right) = l$. We denote $a_j^{t,*} = r^{-1}\left(h^t, l\pi_j^t\right)$, where $r^{-1}\left(h^t, l\right)$ represents the transmission strategy satisfying $r\left(h^t, a_j^t\right) = l$ and $r^{-1}\left(h^t, 0\right) = 0$. Hence, solving the cross-layer optimization in Eq. (1) is equivalent to optimizing the packet scheduling policy $\pi_j^{t,*}$ in the optimal stopping problem, which can be solved using dynamic programming.

---

[7] For simplicity, we consider that the traffic state transition is deterministic once the scheduling action is fixed. The traffic state transition model represents how the traffic state changes over time in the application layer. The deterministic transition can be obtained by transmission strategies at the lower layers, e.g. the retransmission in the MAC layer.
[8] We assume that the channel state transition is independent of the traffic state transition.
[9] The stage corresponds to one time slot. We will interchangeably use state and time slot later in this paper.

## B. Dynamic Programming Solution

In this section, we develop a dynamic programming solution to find the optimal packet scheduling policy $\pi_j$ and the transmission strategy $a_j$. According to the optimality condition of the MDP [13], the optimal packet scheduling $\pi_j^t$ and transmission strategy $a_j^t$ are given as follows:

$$\left(\pi_j^{t,*}, a_j^{t,*}\right) = \arg\max_{\pi_j^t, a_j^t} \left\{ u_j^t\left(s_j^t, \left(\pi_j^t, a_j^t\right)\right) + \alpha \sum_{s_j^t} p\left(s_j^{t+1} \mid s_j^t, \pi_j^t\right) U_j\left(s_j^{t+1}\right) \right\} \quad (2)$$

where $U_j\left(s_j^t\right)$ is the state-value function representing the accumulated net utility from time slot $t$ to $d_j + 1$, which is computed using backward induction as follows:

$$\begin{aligned} & U_j^{d_j+1}\left(s_j^{d_j+1}\right) = 0, \forall s_j^{d_j+1} \\ & U_j^t\left(s_j^t\right) = \max_{\pi_j^t, a_j^t} \left\{ u_j^t\left(s_j^t, \left(\pi_j^t, a_j^t\right)\right) + \alpha \sum_{s_j^t} p\left(s_j^{t+1} \mid s_j^t, \pi_j^t\right) U_j^{t+1}\left(s_j^{t+1}\right) \right\}, \forall s_j^t, t \end{aligned} \quad (3)$$

For the cross-layer optimization for the single-packet transmission, the optimal solutions have the following properties stated in Theorem 1.

**Theorem 1 (Structural properties of the single-packet cross-layer optimization)**:

(i) The optimal packet scheduling policy is a threshold-based policy, i.e.

$$\begin{aligned} & \pi_j^{t,*}\left(\left(1, h^t\right)\right) = \begin{cases} 1 & \text{if } q_j - \lambda \rho^t\left(l, h^t, a_j^{t,*}\right) > \overline{u}_j^t\left(\alpha, h^t\right) \\ 0 & o.w. \end{cases} \\ & \pi_j^{t,*}\left(\left(0, h^t\right)\right) = 0 \end{aligned}$$

where $\overline{u}^t$ is computed by $\overline{u}_j^t\left(\alpha, h^t\right) = \alpha \sum_{h^{t+1}} p\left(h^{t+1} \mid h^t\right) U_j^{t+1}\left(\left(1, h^{t+1}\right)\right)$, and $U_j^t\left(\left(1, h^t\right)\right) = \max\left(q_j - \lambda \rho^t\left(l, h^t, a_j^{t,*}\right), \overline{u}_j^t\left(\alpha, h^t\right)\right)$.

(ii) $\overline{u}_j^t(\alpha, h^t) \geq \overline{u}_j^{t+1}(\alpha, h^t) \geq 0$.

(iii) $\overline{u}_j^t(\alpha_1, h^t) \geq \overline{u}_j^t(\alpha_2, h^t) \geq 0$, if $1 \geq \alpha_1 \geq \alpha_2 \geq 0$.

Proof: See Appendix 1.

From Property (i) in Theorem 1, we note that the optimal packet scheduling policy is determined by comparing the immediate net reward $q_j - \lambda\rho^t\left(h^t, a_j^{t,*}\right)$ to the average future net reward $\overline{u}_j^t(\alpha, h^t)$. The immediate net reward is obtained if the packet is scheduled for transmission at the current time slot, while the average future net reward is the discounted average net reward that the wireless user can obtain if the packet is delayed for future transmission. When $q_j - \lambda\rho^t\left(h^t, a_j^{t,*}\right) > \overline{u}_j^t(\alpha, h^t)$, the wireless user will receive a higher reward if the packet is scheduled for transmission in the current time slot instead of delaying it for future transmission. On the other hand, when $q_j - \lambda\rho^t\left(h^t, a_j^{t,*}\right) \leq \overline{u}_j^t(\alpha, h^t)$, the wireless user prefers to delay the transmission since, on average, a later transmission will lead to a higher reward.

From (ii), we notice that the average future net reward (i.e. the threshold) is decreased as the delay deadline is approached, as shown in Figure 2. This is because, when the delay deadline is far away from the current time, the packet has a higher chance to be transmitted using better channel conditions in the future, and this will result in a lower transmission cost. Hence, the wireless user prefers delaying the transmission by setting a higher threshold for time slots that are further from the deadline.

From (iii), we note that the discount factor can impact the threshold $\overline{u}^t(\alpha, h^t)$: the larger the discount factor is, the higher will the threshold be. There are two extreme cases: one is $\alpha = 0$ and the other one is $\alpha = 1$. When $\alpha = 0$, the threshold $\overline{u}_j^t(\alpha, h^t) = 0, \forall h^t$ and the packet will be scheduled for transmission as early as possible. The packet scheduling with $\alpha = 0$ is referred to as myopic scheduling, since it does not take into account future transmission opportunities. When $\alpha = 1$, the wireless user will delay the packet transmission as much as possible in order to transmit the packet when better channel condition are experienced. One example is illustrated in Figure 2. In this example, when $\alpha = 1$, the packet is transmitted only in the last time slot before it expires (i.e. $\pi_j^* = \left[0\ 0\ 0\ 0\ 1\right]$). When $\alpha = 0.7$, the packet is transmitted in time slot 4, and when $\alpha = 0.3$, the packet is transmitted in the first time slot. The discount factor can be reinterpreted as the risk factor[10]. A high $\alpha$ means that the wireless user is risk-loving, since it prefers delaying the packet for future transmission in order to transmit it during better channel conditions. A low $\alpha$ means that the wireless user is risk-averse since it prefers transmitting the packet as early as possible.

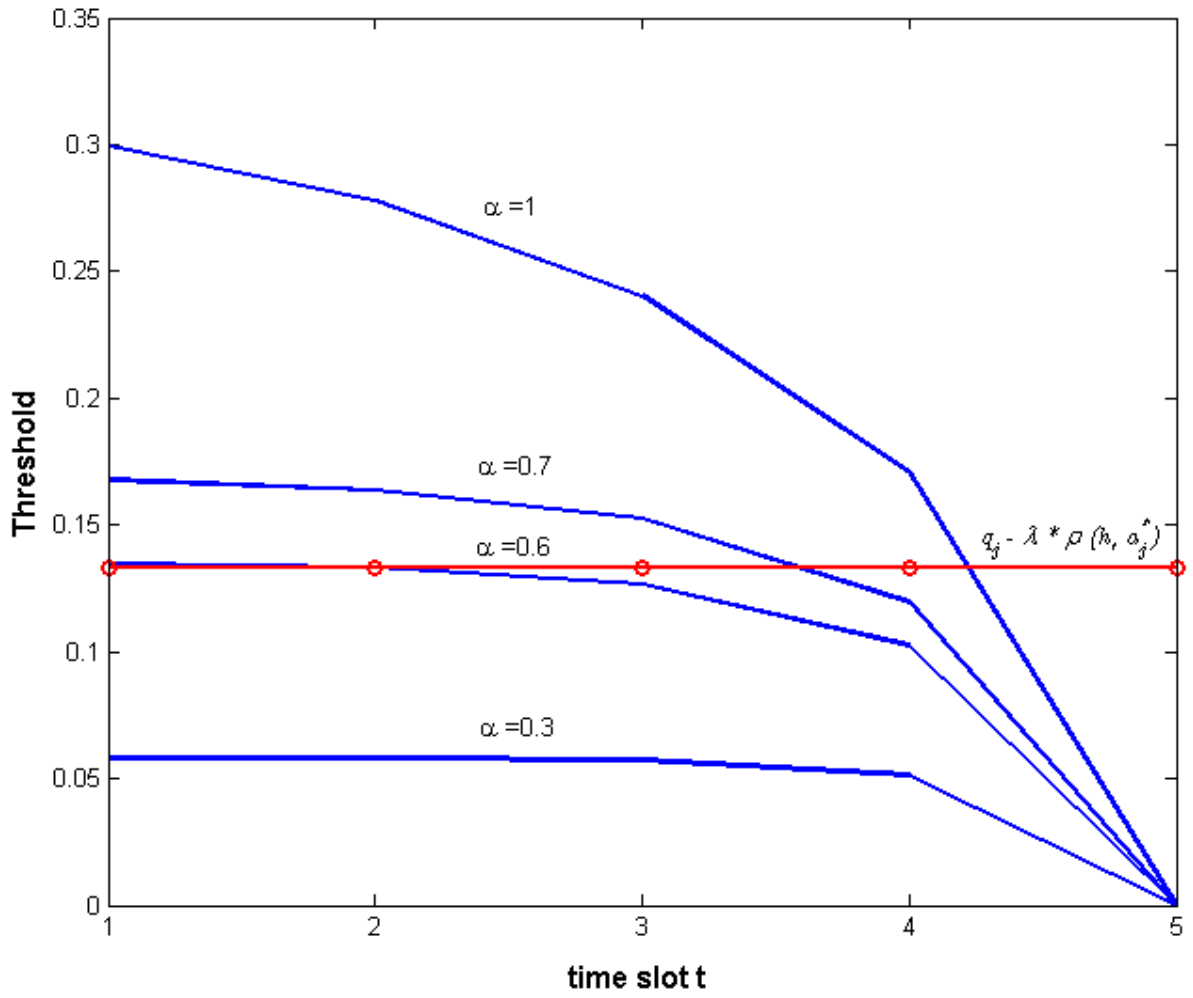

Figure 2. Impact of discount factor $\alpha$ on the scheduling policy

## IV. MDP FOR INDEPENDENTLY DECODABLE PACKETS

In this section, we consider the cross-layer optimization for a group of packets which can be independently decoded. We will delay the discussion for the interdependent packets to Section V.

We consider that there are $N$ packets for transmission, which are independently decodable. Each packet $j \in \{1, \cdots, N\}$ is available for transmission from the time slot $t_j$ to $d_j$. Let $d^{\max} = \max_{1 \le j \le N} \{d_j\}$. We define the traffic state as $B^t = \{b_j^t\}_{j: t_j \le t \le d_j}$, where $b_j^t$ is defined as in Section III. The traffic state $B^t$ includes the traffic states of all packets that are available for transmission (i.e. $t_j \le t \le d_j$) at the current time slot $t$. Accordingly, the state of the wireless user is defined as $s^t = (B^t, h^t)$ to include the traffic state and channel state. The scheduling policy at time slot $t$ is given by $\pi^t = \{\pi_j^t\}_{j: t_j \le t \le d_j}$. Note that within one time slot, there may be multiple packets to be transmitted based on the current traffic state as well as the channel state. The transmission strategy at time slot $t$ is given by

[10] The risk concept is also introduced in [21] to represent the packet loss probability.

$a^t$. The cross-layer action is denoted by $\sigma^t = (\pi^t, a^t)$. The transmission cost is given by $\rho^t\left(l\sum_{j,t_j \le t \le d_j} \pi_j^t, h^t, a^t\right)$, where $\rho^t(\cdot)$ depends on the bitstream length of the transmitted packets, the current channel condition and the current transmission strategy. Then, the immediate net reward at each time slot is given by

$$u^t(B^t, \sigma^t) = \sum_{j:t_j \le t \le d_j} q_j \pi_j^t - \lambda \rho^t\left(l\sum_{j,t_j \le t \le d_j} \pi_j^t, h^t, a^t\right).$$

The difference between the immediate net reward above and the one for the single packet is that the transmission cost $\rho^t$ is a function of the entire bitstream length of all packets to be transmitted at the current time. The cross-layer action is $(\pi^t, a^t)$. (Examples of the cross-layer actions are given in Examples 1 and 2.) An example of the state transition is illustrated in Figure 3.

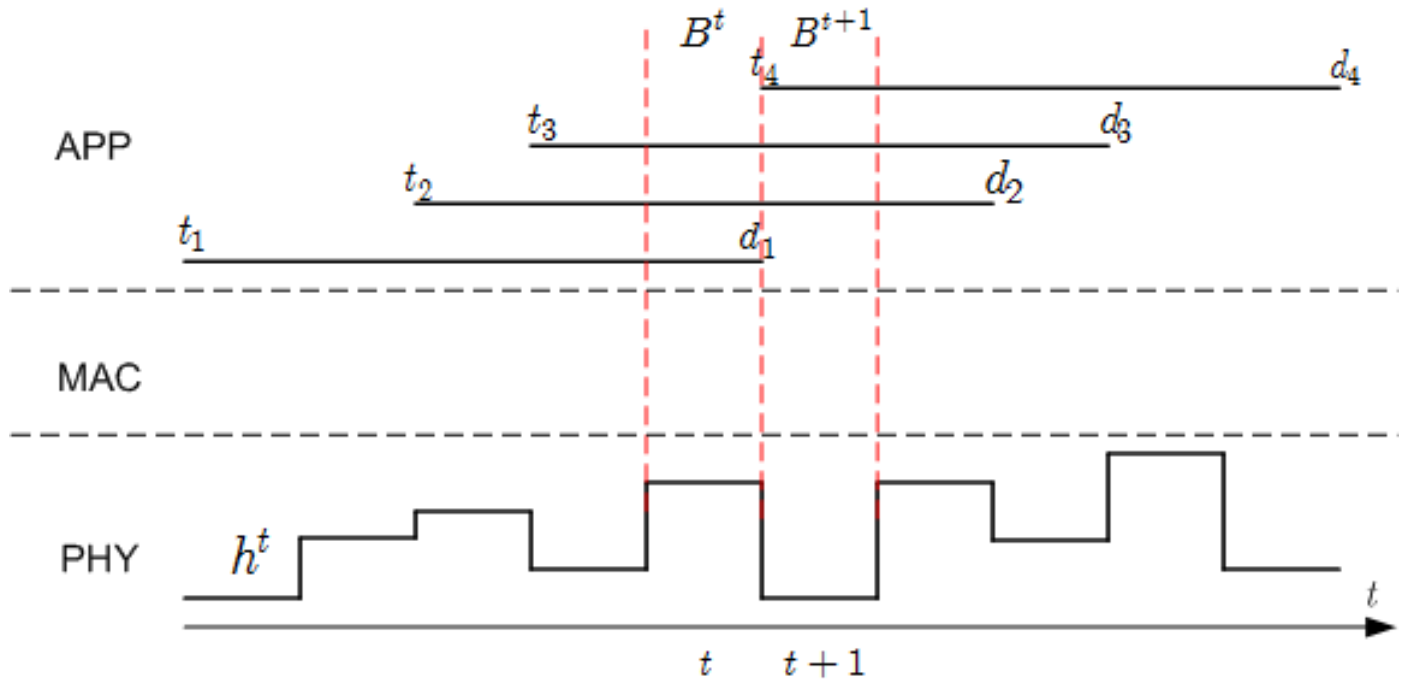

Figure 3. Typical example of state transition (including the traffic state and channel state)

The state transition probability is given by

$$p(s^{t+1} \mid s^t, \pi^t) = \prod_{j,t_j \le t \le d_j} p(b_j^{t+1} \mid b_j^t, \pi_j^t) p(h' \mid h),$$

where $p(b_j^{t+1} \mid b_j^t, \pi_j^t)$ is computed as in Section III.B. Note that the packet with $t_j = t+1$ will be considered in the next time slot with probability 1 and the packets with $d_j = t$ will be discarded in the next time slot with probability 1. The objective in this cross-layer optimization is to maximize the accumulated discounted net utility for all the packets, which is presented as follows:

$$\max_{\pi_j^t \in \{0,1\}, a_j^t \in \mathcal{A}, \forall j,t} \sum_{t=0}^{d^{\max}} \alpha^t u^t\left(B^t, \sigma^t\right)$$

$$s.t. \sum_{t=t_j}^{d_j} \pi_j^t \le 1, \forall j \tag{4}$$

$$r\left(h^t, a^t\right) = l \sum_{j, t_j \le t \le d_j} \pi_j^t, \forall t.$$

As before, the first constraint in Eq. (4) means that each packet is scheduled for transmission within one time slot. The second constraint means that, for those packet scheduled for transmission at time slot $t$, i.e. $\pi_j^t = 1$, the transmission strategy $a^t$ is selected such that the total number of successfully transmission bits equals the bitstream length of the transmitted packets. Note that the packet scheduling $\pi^t$ and transmission strategy $a^t$ should be jointly optimized in order to maximize the net utility.

**Lemma 2**: The cross-layer optimization in Eq. (4) for the $N$ independently decodable packets is a $(d^{\max} + 1)$-stage MDP.

The proof is straightforward based on the above discussion and it is omitted here due to space limitations. Similar to the single-packet transmission, by formulating the cross-layer optimization problem in Eq. (4) as an MDP, the constraint on the packet scheduling is automatically satisfied. For those packet scheduled for transmission at time slot $t$, i.e. $\pi_j^t = 1$, the optimal transmission strategy $a^t$ is selected such that $r\left(h^t, a^{t,*}\right) = l \sum_{j, t_j \le t \le d_j} \pi_j^t$. We denote that $a_j^{t,*} = r^{-1}\left(h^t, l \sum_{j, t_j \le t \le d_j} \pi_j^t\right)$. Hence, solving the cross-layer optimization in Eq. (4) is equivalent to optimizing the packet scheduling policy $\pi^{t,*}$ in the MDP, which can be solved using dynamic programming, which is given as follows:

$$\sigma^{t,*} = \left(\pi^{t,*}, a^{t,*}\right) = \arg\max_{\pi^t, a^t} \left\{ u^t\left(s^t, \sigma^t\right) + \alpha \sum_{s^t} p\left(s^{t+1} \mid s^t, \pi^t\right) U\left(s^{t+1}\right) \right\}, \tag{5}$$

where $U\left(s^t\right)$ is the state-value function representing the accumulated net utility from time slot $t$ to $d_{\max} + 1$. This can be computed using backward induction as follows:

$$U^{d_{\max}+1}\left(s^{d_{\max}+1}\right) = 0, \forall s^{d_j+1}$$
$$U^t\left(s^t\right) = \max_{\pi^t, a^t}\left\{u^t\left(s^t, \left(\pi^t, a^t\right)\right) + \alpha \sum_{s^t} p\left(s^{t+1} \mid s^t, \pi^t\right) U^{t+1}\left(s^{t+1}\right)\right\}, \forall s^t, t \quad (6)$$

However, directly solving the optimization problems in Eqs. (5) and (6) has the following drawbacks:

(i) There are totally $2^N |\mathcal{H}|$ states. It has to check $2^N$ scheduling actions at each state. Hence, the complexity is exponentially increasing with the number of packets to be transmitted and the computation is intractable;

(ii) It disregards the fact that the packets have different transmission priorities. Using the transmission priorities, we are able to develop key structural properties for the dynamic programming solutions as shown in Section C, which provide more insights into how to systematically perform cross-layer optimization for media transmission.

## A. Transmission priority

In Section III, we have already answered the question: at what time should a packet be transmitted when performing single-packet cross-layer optimization? In the multi-packet cross-layer optimization, we are interested in the question: which packet should be transmitted first? In order to determine the transmission orders for the packets, we first define the marginal utility for packet $j$ at time slot $t$ as follows:

$$\Delta u_j^t\left(s^t, \sigma_{-j}^t\right) = \begin{cases} 0 & \text{if } b_j^t = 0 \\ q_j - \lambda \rho^t\left(s^t, \sigma_{-j}^t\right) - \bar{u}_j^t\left(s^t, \sigma_{-j}^t\right) & o.w. \end{cases} \quad (7)$$

where $\sigma_{-j}^t$ is the cross-layer action for the packets except packet $j$, $\bar{u}_j^t\left(s^t, \sigma_{-j}^t\right)$ is the net utility the wireless user can obtain if it delays the packet transmission, serving as a threshold used to determine whether packet $j$ is scheduled for transmission or not, which is similar to the threshold for single-packet transmission. $q_j - \lambda \rho^t\left(s^t, \sigma_{-j}^t\right)$ is the net utility if packet $j$ is scheduled at time slot $t$. Note that $\bar{u}_j^t\left(s^t, \sigma_{-j}^t\right)$ and the transmission cost $\rho^t\left(s^t, \sigma_{-j}^t\right)$ may depend on the cross-layer actions of other packets that can be transmitted at the current time slot. The specific computation is presented in Sections B and C. Then, the marginal utility $\Delta u_j^t\left(s^t, \sigma_{-j}^t\right)$ is the amount by which the utility can be increased if the packet is transmitted at the current time slot $t$ rather than delaying it for future transmission. From Section III.B, we know that, for the single-packet transmission, if $\Delta u_j^t(s^t, \sigma_{-j}^t) = \Delta u_j^t((1, h^t)) > 0$, then the packet is scheduled for transmission. Using the marginal utility, we are able to formally define the transmission priorities between packets as follows.

**Definition (Transmission Priority)**: Packet $j$ has a higher transmission priority than packet $k$ (denoted by $j \lhd k$) at time slot $t$ if $\Delta u_j^t(s^t, [\sigma_j^t, \varnothing_k, \sigma_{-j,-k}^t]) \geq \Delta u_k^t(s^t, [\varnothing_j, \sigma_k^t, \sigma_{-j,-k}^t])$ for $\forall s^t$ and $\sigma_{-j,-k}^t$ and where $\varnothing_k$ means that packet $k$ is not transmitted and $\sigma_{-j,-k}^t$ is the cross-layer action of other packets, except packets $j$ and $k$.

The transmission priority defined above indicates that, if both packets $j$ and $k$ are available for transmission (i.e. $b_j^t = b_k^t = 1$) at time slot $t$, then packet $j$ will be transmitted before packet $k$ under any channel conditions. Note that it is possible that both packets can be transmitted at time slot $t$. However, if the lower priority packet $k$ is available for transmission before the higher priority packet $j$ (i.e. $t_k < t_j$), then it is possible that packet $k$ can be transmitted earlier than packet $j$. We will discuss this in Section C. This priority definition can be easily extended to multiple packets. However, in order to determine the transmission priorities between packets directly using the above definition, we have to compare the marginal utilities under all the possible states as well as other packets' cross-layer actions. This can only be done after solving the cross-layer optimization. In Sections B, C and V, we will determine the priorities between packets only based on the delay deadlines, distortion impacts and the dependencies of both packets without solving the cross-layer optimization. Note that the transmission priority considered here is different from the simplified priority definition which only depends on the distortion impact [6].

## B. Cross-layer optimization with linear transmission cost

First, we consider that the transmission cost $\rho^t$ is a linear function of the total bitstream length of the packets to be transmitted which corresponds to the case that packets are not self-congested as in [7], as shown in Example 1, i.e.

$$\rho^t\left(\sum_{j:t_j \leq t \leq d_j} \pi_j^t l, h^t, a^t\right) = \sum_{j:t_j \leq t \leq d_j} \pi_j^t \rho^t\left(l, h^t, a_j^t\right) \tag{8}$$

By having a linear transmission cost, the dynamic programming solution[11] to the $N$ -packet cross-layer optimization in Eq. (4) can be decomposed into $N$ single-packet cross-layer optimization each of which is solved by the dynamic programming solution developed in Section III.B, as shown in the following theorem.

**Theorem 2 (Structural properties of cross-layer optimization for multiple independently decodable packets with linear transmission cost)**: The cross-layer optimization for $N$ independently decodable packets can be decomposed into $N$ single-packet cross-layer optimization problems each of which is a $(d_j - t_j + 1)$ -stage Markov decision process solved using the dynamic programming developed in Section III.B. In other words, $\pi^{t,*} = \{\pi_j^{t,*}\}_{j:t_j \le t \le d_j}, a^{t,*} = \{a_j^{t,*}\}_{j:t_j \le t \le d_j}$ and $U^t(s^t) = \sum_{j:t_j \le t \le d_j} U_j^t(s_j^t)$.

Proof: see Appendix 2.

By decomposing the cross-layer optimization of the $N$ independently decodable packets with linear transmission cost, the complexity is linearly increased with the number of packets.

Based on the transmission priority definition in Section A, we have the following lemma:

**Lemma 3**: For packets $j$ and $k$ with linear transmission cost, if $q_j \ge q_k$ and $d_j \le d_k$ , then $\Delta u_j^t((1,h^t)) \ge \Delta u_k^t((1,h^t)), \forall h^t$, $\max\{t_j, t_k\} \le t \le d_j$, i.e. $j \lhd k$ .

Proof: see Appendix 3.

Lemma 3 indicates that if $q_j \ge q_k$ and $d_j \le d_k$, the marginal utility of packet $j$ is larger than that of packet $k$ at any network condition at time slot $\max\{t_j, t_k\} \le t \le d_j$ in which both packets are available for transmission, no matter what cross-layer actions are taken for other packets. We only use the distortion impacts and delay deadlines to determine the transmission priority before performing the cross-layer optimization. The priority definition also plays an important role in the case when the transmission cost and transmission rate are not linear, which is discussed in the next section.

---

[11] In this solution, we do not consider the constraint that the total amount of transmission time available at each time slot should be less than the length of time slot. By considering this solution, the optimal cross-layer optimization can solved using the method proposed in Section IV.C.

## C. Cross-layer optimization with convex transmission cost

In this section, we consider a more general scenario in which the transmission cost $\rho^t\left(l\sum_{j:t_j\le t\le d_j}\pi_j^t, h^t, a^t\right)$ is a convex and increasing function of the total bitstream length of the transmitted packets, i.e. $l\sum_{j:t_j\le t\le d_j}\pi_j^t$, which corresponds to the case that the packets are self-congested, as shown in Example 2. In this case, since $\rho^t\left(l\sum_{j:t_j\le t\le d_j}\pi_j^t, h^t, a^t\right)$ is not a linear function, the cross-layer optimization for the $N$-independently decodable packets is not decomposable. However, similar to Lemma 2, we can compare the transmission priorities of different packets with convex transmission cost.

**Lemma 4**: For packets $j$ and $k$ with convex transmission cost, if $q_j \ge q_k$ and $d_j \le d_k$, then $j \lhd k$.

The proof is similar to the one for Lemma 3, but with convex transmission cost.

It is worth to note that we cannot prioritize packets $j$ and $k$ when $q_j > q_k$ and $d_j > d_k$ since it is difficult to determine the marginal utility for each packet without solving the cross-layer optimization.

*C.1 Priority graph expression for traffic state*

Given $M(M \le N)$ packets with arbitrary distortion impact $q_j$, arrival time $t_j$ and delay deadline $d_j$, $j \in \{1,\cdots,N\}$, which are available for transmission at the same time, we can construct a priority graph based on the transmission priorities, which is a DAG $PG = \langle V_G, E_G \rangle$ as shown in Figure 4. The node set $V_G \subseteq \{1,\cdots,N\}$ represents the set of packets. Each element of the directed edge set $E_G$ represents a pair $(j,k)$ meaning that $j \lhd k$ and there is no other packet $j'$ such that $j \lhd j' \lhd k$. Any path directed from $k$ to $j$ means that $j \lhd k$. Any root of the priority graph $PG$ is a packet whose priority is not less than any other packets. A priority graph may have multiple roots, e.g. the graph in Figure 4 (c), if the packets cannot be prioritized. The priority graph compactly expresses the transmission priorities between packets, which can be used to represent the traffic state at each time slot as shown in Figure 5.

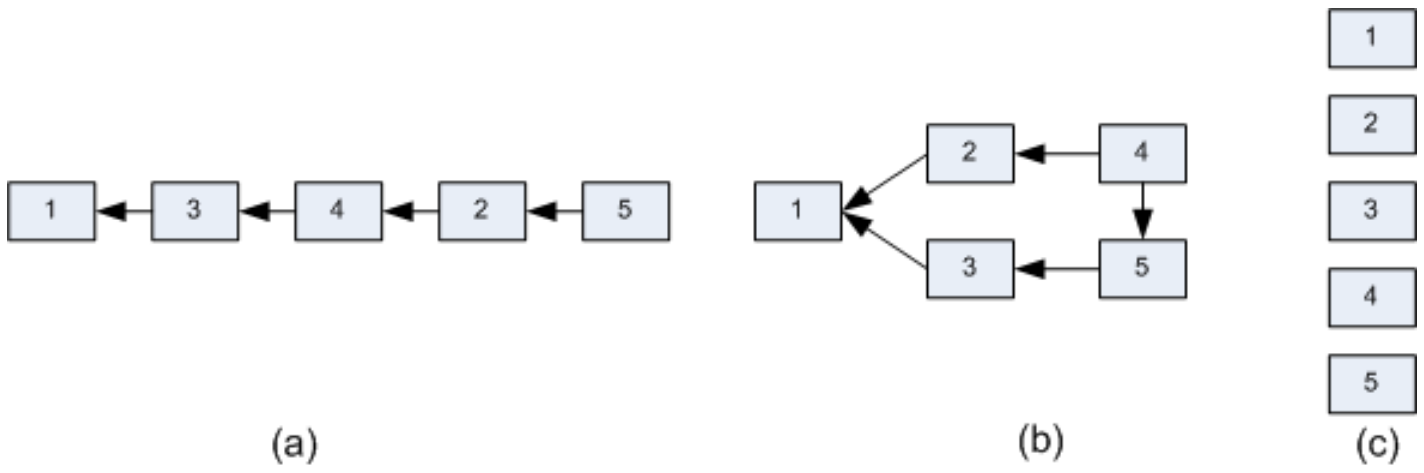

Figure 4. Typical priority graphs with five packets: (a) all packets are prioritized (chain); (b)only part of the packets are prioritized (e.g. the priorities of packets 2 and 3 are unknown); (c) no packets can be prioritized.

Specifically, for each traffic state $B^t$ at time slot $t$, the set of packets which have not been transmitted is $J^t(B^t) = \{j : b_j^t = 1, t_j \leq t \leq d_j\}$. Then, we can construct the priority graph $PG(B^t)$ by taking $J^t(B^t)$ as the nodes and the transmission priorities between the packets as the directed edges. It is clear that the traffic state can be uniquely represented by the priority graph. We interchangeably use the terms priority graph and traffic state in the remainder of the paper.

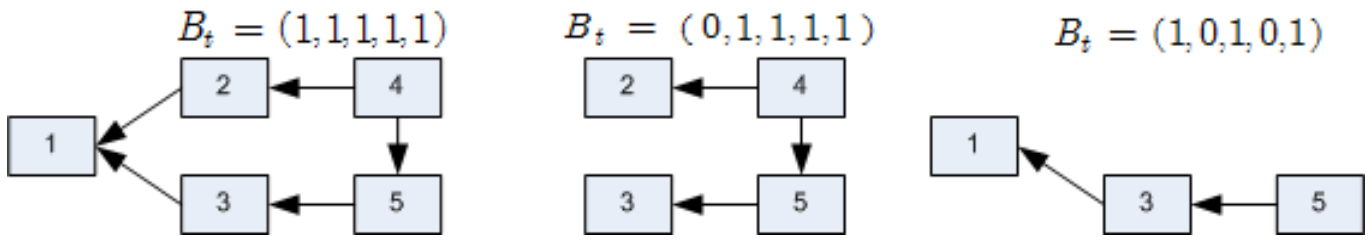

Figure 5. Traffic states with five packets and their corresponding priority graphs

*C.2 Cross-layer optimization as travelling state-tree*

In the following, we will present the cross-layer solution derived based on the priority graph. First, we present here the traffic state transition expressed by the priority graph. In state $s^t = (B^t, h^t)$, the traffic state is expressed by the priority graph $PG(B^t)$. The cross-layer action $\sigma^t = (\pi^t, a^t)$ is deployed, which leads to packets $J_{TX}^t(s^t, \sigma^t) = \{j : \pi_j^t = 1, j \in J^t(B^t)\}$ to be transmitted. The set of the packets which will be expired at time slot $t$ (i.e. $d_j = t$) is denoted by $J_{EXP}^t = \{j : d_j = t\}$. We define a post-state as $\tilde{s}^t = (\tilde{B}^t, h^t)$, where $\tilde{B}^t$ is the post traffic state after the transmission and each element $b_j^t$ of $\tilde{B}^t$ is 1 if $j \in J^t(B^t)/(J_{TX}^t(B^t, \pi^t, a^t) \cup J_{EXP}^t)$[12] and otherwise, is 0. In the other words, the post traffic state represents those packets that have not been transmitted at the current time slot and can be transmitted at next time slot. Then, the post traffic state $\tilde{B}^t$ deterministically transits to the next traffic state $B^{t+1}$ by adding the new arriving packets (with $t_j = t+1$), as shown in Figure 6.

The channel state $h^t$ transits to $h^{t+1}$ with the transition probability $p(h^{t+1} \mid h^t)$. Then we are able to define the post-state value function as follows:

$$\bar{u}^t\left(\left(\tilde{B}^t, h^t\right)\right) = \alpha \sum_{h^{t+1}} p\left(h^{t+1} \mid h^t\right) U^{t+1}\left(B^{t+1}, h^{t+1}\right). \tag{9}$$

where the next state $B^{t+1}$ is reached from the post state $\tilde{B}^t$. The post-state value function represents the average future net utility starting from the state $\left(\tilde{B}^t, h^t\right)$. Note that $\bar{u}^t\left(\left(\tilde{B}^t, h^t\right)\right)$ is the same as $\bar{u}^t\left(\alpha, h^t\right)$ defined in Theorem 1 if $N = 1$.

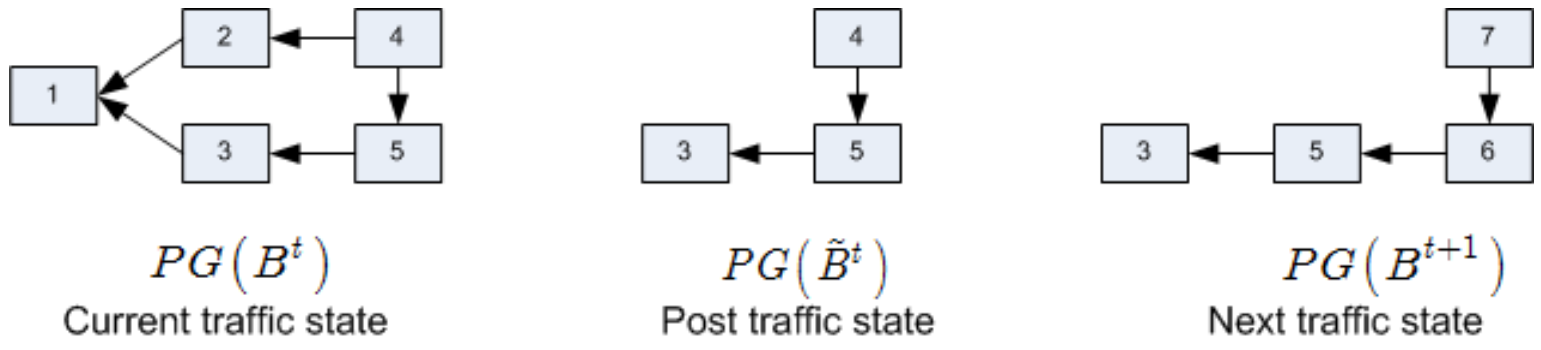

Figure 6. Example of traffic state transition through the post traffic state: The post traffic state is obtained by deleting the transmitted packets from the current traffic state and next traffic state is obtained by adding the new arriving packets (with $t_j = t+1$).

Next, we discuss how the optimal cross-layer solution can be found for each state $s^t$ with the traffic state $PG(B^t)$. We define an operation $root(PG)$ to extract all the roots existing in $PG$. For each priority graph $PG$, we are able to induce a state tree $T(PG) = \langle V_T, E_T \rangle$, where $V_T$ is the set of nodes each of which represents a priority graph and $E_T$ is the set of directed edges each of which represents a pair $(PG', PG'')$ and $PG''$ is the child of $PG'$. A state tree can be induced from a given priority graph $PG$ as follows: the root of the state tree is $PG$ and the children of any node $PG'$ are the graphs $PG'' = PG' / \{v\}$, where $v \in root(PG')$. In other words, $PG''$ is obtained by deleting one of the roots in the priority graph $PG'$ and corresponding edges directed to that root. If $PG' = \varnothing$, it is the leaf of the state tree. The state tree is constructed as in Algorithm 1 in Appendix 5. The examples of the state tree are illustrated in Figure 7.

---

[12] Here $J / J_{TX} = J - J \cap J_{TX}$.

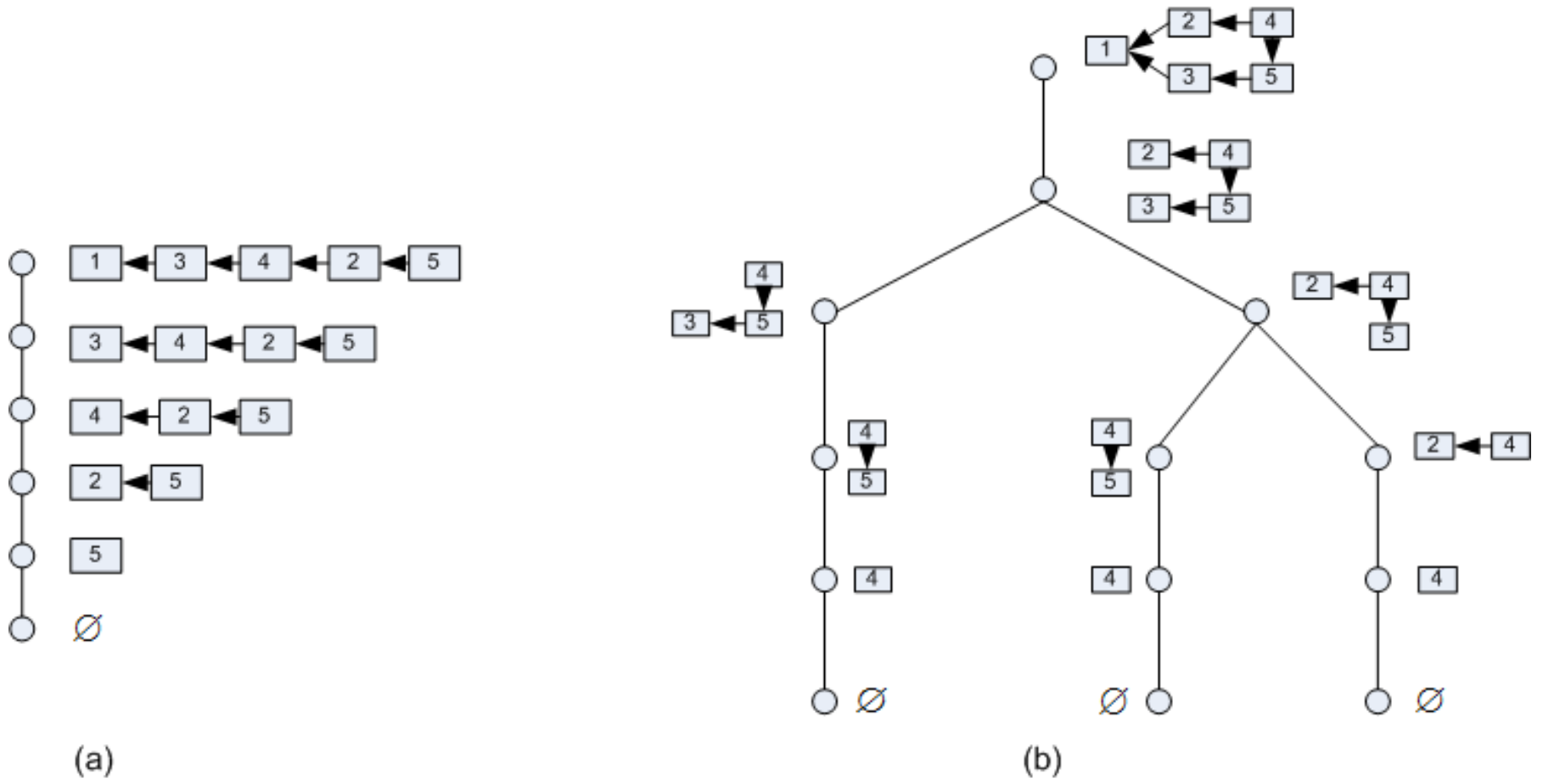

Figure 7. State trees induced from (a) the graph in which all packets are prioritized (chain) (b) the graph in which some packets are prioritized and others are not.

We denote by $T(B^t) = (V_T(B^t), E_T(B^t))$ the state tree induced from the priority graph $PG(B^t)$ which represents the traffic state $B^t$. The following theorem presents the optimal cross-layer solution at any state $s^t = (B^t, h^t)$.

**Theorem 3 (Structural properties of cross-layer optimization for multiple independently decodable packets with convex transmission cost)**:

(i) The optimal packet scheduling policy in state $s^t = (B^t, h^t)$ can be computed by repeating the following two phases:

Phase 1: Select the packet with the highest marginal utility from the root of priority graph $PG^k$ to transmit:

$$j_k = \arg \max_{j \in root(PG^k)} \{\Delta \overline{u}_j^t\}$$

Phase 2: Determine whether the best packet should be transmitted or not:

$$\pi_{j_k}^t = \begin{cases} 1 & if \ \Delta \overline{u}_{j_k}^t((PG^k, h^t)) > 0 \\ 0 & o.w. \end{cases}$$

where the marginal utility is given by $\Delta \overline{u}_{j_k}^t(PG^k, h^t) = q_{j_k} - \lambda[\rho(kl, h^t) - \rho((k-1)l, h^t)] + \overline{u}^t((PG^k / \{j_k\}, h^t)) - \overline{u}^t((PG^k, h^t))$ and $\overline{u}^t(\tilde{s}^t)$ is the post-state value function for the post state $\tilde{s}^t$ computed as in (9), and $PG^k$ is computed by $PG^k = PG^{k-1} / \{j_{k-1}\}$ and

$PG^0 = PG(B^t)$.

(ii) The state-value function is updated by

$$U^t((B^t,h^t)) = \sum_{k=1}^{k_{\max}} q_{j_k} - \lambda\rho(k_{\max}l,h^t) + \overline{u}^t((PG^{k_{\max}},h^t))$$
$$U^t((\mathbf{0},h^t)) = 0$$

where $k_{\max}$ is the maximum $k$ such that

$$\{q_{j_k} - \lambda[\rho(kl,h^t) - \rho((k-1)l,h^t)] + \overline{u}^t((PG^k/\{j_k\},h^t))\} > \overline{u}^t((PG^k,h^t))$$

Proof: see Appendix 4.

Property (i) in Theorem 3 can be easily explained using the traffic state tree. Starting from the root $PG(B^t)$ of the traffic state tree $T(B^t)$, in Phase 1, we compare all the children of $PG(B^t)$ and select the best child $PG(B^t)/\{j_1\}$ (i.e. select packet $j_1$ to transmit) which gives the highest marginal utility. In Phase 2, we determine whether packet $j_1$ is transmitted or not by comparing the marginal utility with 0. The marginal utility is defined similar to Eq. (7) but it depends on the transmission of the packets with higher priorities. The search procedure in property (i) is referred to as the "*travelling state tree*" algorithm in which Phase 1 determines which child in the tree should be reached and Phase 2 determines whether the travel should be stopped or not. We further note that, to determine the optimal cross-layer action at the current state $s^t$, we have to know the post-state value function for all the states that are actually the successors of the traffic state $B^t$ in the state tree $T(B^t)$. In next section, we will quantify the complexity of the cross-layer optimization.

*C.3 Complexity of cross-layer optimization*

From Theorem 3, we know that the complexity of the cross-layer optimization is determined by the number of states (corresponding to the computation complexity) that need to perform the cross-layer optimization, and the number of post-states (corresponding to the storage overhead) whose value functions need to be stored. From the state transition presented in Section C.2, we note that the number of visited states and the number of post-states have the following relationship.

**Lemma 5**. The number of post-states at time slot $t$ is the same as the number of states visited at time slot $t+1$.

*Proof*: This can be easily proved since the packets with $t_j = t+1$ are deterministically available for transmission. We omit the details here in order to save the space.■

We know that, when the standard dynamic programming shown in Eqs. (5) and (6) is performed, we have to perform the cross-layer optimization for all the possible states which is exponentially increasing with the number of packets to be transmitted. However, since the transmission starts with the initial traffic state $B^0 = \mathbf{1}$ with each element being 1, some traffic states will never be visited. Hence, we do not need to compute the optimal cross-layer action for those states. In the below, we will examine how many states can be visited at each time slot if the initial traffic state is $B^0 = \mathbf{1}$.

It is easy to know that, for any two packets $j,k$, if $t_j < t_k$, then it is possible that $b_j^t = 1$ while $b_k^t = 0$ regardless of the priorities of packets $j$ and $k$. Next, we have the following lemma which states the relationship among $b_j^t$ and $b_k^t$.

**Lemma 6**: If $t_j \le t_k$ and $j \lhd k$, then any traffic state $B^t\left(t_k \le t \le \min\left(d_j, d_k\right)\right)$ must not have $b_j^t = 1, b_k^t = 0$.

*Proof*: In time slot $t_k$, it is only possible that $b_j^{t_k} = b_k^{t_k} = 1$ or $b_j^{t_k} = 0, b_k^{t_k} = 1$. Starting from time slot $t_k$, we can only obtain $b_j = b_k = 1$, or $b_j = 0, b_k = 1$ or $b_j = b_k = 0$ since $j \lhd k$. ■

All the packets with $t_j \le t \le d_j$ can be potentially transmitted in time slot $t$. However, as shown in Lemma 4, some traffic states cannot be visited. To show which state can be visited at time slot $t$, we construct a new DAG graph $\widehat{PG}^t = \left(\widehat{V}_G, \widehat{E}_G\right)$ where $\widehat{V}_G = \{j : t_j < t \le d_j\}$ and $\widehat{E}_G = \{(j,k) \mid t_j \le t_k, j \lhd k\}$. This graph is different from the priority graph constructed for each traffic state as follows: the edges in the priority graph represent the transmission priorities between packets while the edges in this new DAG graph not only satisfy the transmission priorities but also meets the arrival time constraints. Note that the DAG $\widehat{PG}^t$ does not include the packets that arrive at the current time slot $t$. We can also induce a state tree $\widehat{T}\left(\widehat{PG}^t\right) = \left(\widehat{V}_{\widehat{T}}, \widehat{E}_{\widehat{T}}\right)$ from the graph $\widehat{PG}^t$ using Algorithm 1. Then, the node in the graph $\widehat{PG}^t$ represents the remaining packets that have not been transmitted before time slot $t$, which is the post-state at time slot $t-1$. The packets to be transmitted at time slot $t$ are the packets that have not been transmitted before time slot $t$ and the packets that arrive during the current time slot. It is clear that the traffic state in the state tree $\widehat{T}\left(\widehat{PG}^{t}\right)$ satisfies the constraint illustrated in Lemma 6. From Lemma 5, we know that the number of traffic states that can be visited at time slot $t$ is equal to the number of post-traffic states in time slot $t-1$, which is the number of distinct nodes in the state tree $\widehat{T}\left(\widehat{PG}^{t}\right)$. To quantify the complexity of the cross-layer optimization, we define the disconnection degree of a DAG as follows:

**Definition 2 (Disconnection degree)**: The disconnection degree $\phi(PG)$ of a DAG $PG$ is the number of packet pairs for each of which there exists no path connecting these two packets.

For the priority graph, the disconnection degree represents the number of packet pairs that cannot be prioritized. For example, the disconnection degrees in the priority graphs in Figure 4 are 0, 2 and 10, respectively. The number of the distinct nodes (except $\varnothing$) in the state tree $T(B)$ is given by $N(B)+\phi(PG(B))$, where $N(B)$ is the number of packets in the priority graph. The number of distinct nodes in the state trees shown in Figure 7 is 5+0 and 5+2, respectively. It is worth to note that the disconnection degree of the priority graph is determined by the characteristics of the media packets (specifically the delay deadlines and distortion impacts). Then, the number of states that can be visited and the number of post-states at time slot $t$ are given by Corollary 1.

**Corollary 1**: Starting in the initial state $s^{0}=\left(B^{0},h^{0}\right)$, the number of states that can be visited in time slot $t$ is $|\mathcal{H}|\left(N^{t}\left(\widehat{B}^{t}\right)+\phi\left(\widehat{PG}\left(\widehat{B}^{t}\right)\right)\right)$ and the number of the post-states whose value functions need to be stored is given by $|\mathcal{H}|\left(N^{t}\left(\widehat{B}^{t+1}\right)+\phi\left(\widehat{PG}\left(\widehat{B}^{t+1}\right)\right)\right)$.

*Proof*: The proof is straightforward based on the above discussion and it is omitted here due to space limitations.■

From the above analysis, we know that the complexity of the cross-layer optimization is linearly increasing with the disconnection degree of the priority graph, which is determined by the characteristics of the media data. In next section, we will extend the priority graph to the case with interdependent packets and develop the corresponding cross-layer optimization solution.

## V. MDP FOR INTERDEPENDENT PACKETS

In this section, we aim to develop a cross-layer optimization solution for $N$ interdependent packets. As discussed in Section II, the dependency is expressed by a directed acyclic graph which is called a dependency graph (DG). In the following, we first examine the transmission priorities between the interdependent packets.

**Lemma 7**: If $j \prec k$, then $j \lhd k$.

The proof is similar to the one for Lemma 2, but with convex transmission cost.

From Lemma 3, we note that, if packet $k$ depends on packet $j$, then packet $j$ will be transmitted earlier than packet $k$. It is obvious since, in order to decode packet $k$, packet $j$ must be available at the destination. We further note that, after introducing the interdependency, the packets $j$ and $k$ satisfy $q_j \geq q_k$ and $d_j \leq d_k$, packet $j$ may not have a higher priority than packet $k$. However, we have the following lemma that enables us to compare the priorities of different packets.

**Lemma 8**: If $q_j \geq q_k$ and $d_j \leq d_k$, and $\{k' : k \prec k'\} \subseteq \{j' : j \prec j'\}$, then $j \lhd k$.

The proof is straightforward based on the above discussion and it is omitted here due to space limitations.

Lemma 6 tells us that if two packets are not connected in the DG, the transmission priority between these two packets can be determined by their distortion impacts, delay deadlines and the descendents that depend on them. For interdependent packets, besides the traffic state $B^t$ which represents whether the packets $\{j : t_j \leq t \leq d_j\}$ have been transmitted or not, we further define a dependency state $D^t = \{b_k^t \mid k \in K^t\}$ where $K^t = \left\{k : d_k < t, \exists j \in J^t, j \text{ directely depends on } k\right\}$. $K^t$ is the set of packets that expire before time slot $t$ and on which the current packets $J^t$ directly depend. Then $D^t$ represents the traffic states of the expired packets $K^t$. The state at time slot $t$ is given by $s^t = (D^t, B^t, h^t)$. The transition of the traffic state $B^t$ is the same as with independently decodable packets. The update of the dependency state $D^t$ is performed by deleting the packets on which no packets in next time slot depend on and adding the packets expired in the current time slot and on which future packets will depend (note that $D^t$ represents the traffic states for those packets which have expired). Similarly, we can compute the post state $\tilde{s}^t = \left(D^t, \tilde{B}^t, h^t\right)$, which is the state after the transmission.

Similar to the priority graph for independently decodable packets, we can construct a priority graph $PG((D^t, B^t))$ for the state $s^t = (D^t, B^t, h^t)$, in which the nodes are the packets $\left\{j : j \in J^t; b_j^t = 1; b_k^t = 0 \text{ if } \exists k \in K^t \text{ such that. } k \prec j\right\}$. In other words, the nodes in the priority graph $PG\left(\left(D^t, B^t\right)\right)$ are those packets that are decodable (i.e. whose ancestors have been transmitted) but which have not yet been transmitted. It is clear that the priority graph $PG\left(\left(D^t, B^t\right)\right)$ is constructed based on both the traffic state and dependency state. Then, similar to the state tree construction for the independently decodable packets, we can induce the state tree $T\left(D^t, B^t\right)$ from the priority graph $PG\left(D^t, B^t\right)$. Similar to Theorem 4, the cross-layer optimization for the state $s^t = \left(D^t, B^t, h^t\right)$ is performed as illustrated in Theorem 5.

**Theorem 4 (Structural properties of cross-layer optimization for multiple interdependent packets with partial priorities)**.

(i) The optimal packet scheduling policy in state $s^t = \left(D^t, B^t, h^t\right)$ can be computed by repeating the following two phases.

Phase 1: Select the packet with the highest marginal utility from the roots of priority graph $PG^k$ to transmit:

$$j_k = \arg \max_{j \in root\left(PG^k\right)} \left\{\Delta \overline{u}_j^t\left(PG^k, h^t\right)\right\}$$

Phase 2: Determine whether the best packet to be transmitted or not:

$$\pi_{j_k}^t = \begin{cases} 1 & if \ \Delta \overline{u}_{j_k}^t\left(\left(PG^k, h^t\right)\right) > 0 \\ 0 & o.w. \end{cases}$$

where the marginal utility is given by $\Delta \overline{u}_{j_k}^t\left(PG^k, h^t\right) = q_{j_k} - \lambda\left[\rho\left(kl, h^t\right) - \rho\left((k-1)l, h^t\right)\right] + \overline{u}^t\left(\left(PG^k / \{j_k\}, h^t\right)\right) - \overline{u}^t\left(\left(PG^k, h^t\right)\right)$ and $\overline{u}^t\left(\tilde{s}^t\right)$ is the post-state value function for the post state $\tilde{s}^t$ computed as in (9), and $PG^k$ is computed by $PG^k = PG^{k-1} / \{j_{k-1}\}$ and $PG^0 = PG\left(B^t\right)$.

(ii) The state-value function is updated by

$$\begin{aligned} & U^t\left(\left(D^t, B^t, h^t\right)\right) = \sum_{k=1}^{k_{\max}} q_{j_k} - \lambda \rho\left(k_{\max} l, h^t\right) + \overline{u}^t\left(\left(PG^{k_{\max}}, h^t\right)\right) \\ & U^t\left(\left(\mathbf{0}, h^t\right)\right) = 0 \end{aligned}$$

where $k_{\max}$ is the maximum $k$ such that

$$\left\{q_{j_k} - \lambda\left[\rho\left(kl, h^t\right) - \rho\left((k-1)l, h^t\right)\right] + \overline{u}^t\left(\left(PG^k / \{j_k\}, h^t\right)\right)\right\} > \overline{u}^t\left(\left(PG^k, h^t\right)\right)$$

Proof: The proof is similar to the one to Theorem 3 and thus, is omitted here due to space limitations.

Similarly, the cross-layer action selection shown in property (i) in Theorem 4 can also be easily explained as "*travelling the traffic state tree*". Starting from the root $PG(D^t,B^t)$ of the traffic state tree $T(D^t,B^t)$, in Phase 1, we compare all the children of $PG(D^t,B^t)$ and select the best child $PG(D^t,B^t)/\{j_1\}$ (i.e. select packet $j_1$ to transmit) which gives the highest marginal utility. In Phase 2, we determine whether packet $j_1$ is transmitted or not by comparing the marginal utility with 0. The marginal utility is defined similar to Eq. (7), but it depends on the transmission of the packets with higher priorities as well as the dependency state. We further note that, to determine the optimal cross-layer action at the current state $s^t$, we have to know the post-state value function for all the states that are actually the successors of the traffic state $B^t$ in the state tree $T(D^t,B^t)$. From Corollary 1, we note that there are $N^t(D^t,B^t)+\phi(PG(D^t,B^t))$ post-state value functions to be stored and there are at most $N^t(D^t,B^t)+\phi(PG(D^t,B^t))$ comparisons in Phase 1 to compute the optimal cross-layer action for the state $s^t=(D^t,B^t,h^t)$, where $N^t(D^t,B^t)=\#\left\{j : j\in J^t; b_j^t=1; b_k^t=0 \text{ if } \exists k\in K^t \text{ such that. } k\prec j\right\}$ is the number of the packets that are decodable and not yet transmitted.

The number of visited states at time slot $t$, when starting from the initial state $B^0=\mathbf{1}$, depends on the dependency state $D^t$ and $B^t$. The number of possible dependency states is $2^{|K_t|}$ while the number of $B^t$ is the same as in Corollary 1 which is $N^t\left(\widehat{B}^t\right)+\phi\left(\widehat{PG}\left(\widehat{B}^t\right)\right)$. Hence, the total number of visited states at time slot $t$ is $2^{|K_t|}\left(N^t\left(\widehat{B}^t\right)+\phi\left(\widehat{PG}\left(\widehat{B}^t\right)\right)\right)$, which is the same as the total number of post-states in time slot $t-1$. From this analysis, we know that the complexity of the cross-layer optimization exponentially increases with the number of dependency states and linearly increases with the disconnection degree of the priority graph. Moreover, it is worth mentioning that the cross-layer optimization for interdependent packets even with linear transmission costs cannot be decomposed due to the interdependencies between packets.

## VI. SIMULATION RESULTS

In this section, we perform several numerical experiments to compare the performance of various state-of-art solutions for multimedia communications with the proposed framework.

## A. Performance comparison of various cross-layer solutions for video transmission

In this section, we compare our proposed cross-layer solution with several start-of art solutions which only consider either the media characteristics or the time-varying channel conditions. In the experiment, to compress the video data, we used a scalable video coding scheme [15], which is attractive for wireless streaming applications because it provides on-the-fly application adaptation to channel conditions, support for a variety of wireless receivers with different resource capabilities and power constraints, and easy prioritization of various coding layers and video packets. We choose for this experiment three video sequences (Foreman, Coastguard and Mobile at CIF resolutions, 30 frames/second), exhibiting different motion activities. The video sequences "Foreman" and "Coastguard" are encoded at the bit rate of 512 kbps and "Mobile" is encoded at 1024kbps. Each Group of Picture (GOP) contains 8 frames and each encoded video frame can tolerant a delay of 266ms corresponding to the duration of GOP. The transmission cost is the amount of power consumed during the packet transmission and is computed as in Example 2. The normalized channel gain varies from 0 to 1 and is modelled as in [16] as a FSMC with 5 states. The cross-layer action includes the packet scheduling at the APP layer and power allocation at the PHY layer. The transmission strategies at the MAC are not considered here. We consider three comparable solutions: (i) our proposed cross-layer solution which takes into account both the heterogeneous multimedia traffic characteristics (e.g. delay deadlines, distortion impacts and dependencies etc.) and time-varying network conditions; (ii) the cross-layer solution [6] which only considers the distortion impact of each media packet and adapts the transmission strategies based on the observed channel conditions; (iii) the solution performing the rate-distortion optimization assuming the constant channel conditions (i.e. using average channel conditions) as in RaDiO [7].

In Figure 8 (a)~(c), , we show the Peak-Signal-to-Noise Ratio (PSNR)- (normalized) energy curves under different transmission solutions for the three video sequences. From these figures, we note that our proposed cross-layer optimization solution outperforms both the conventional "distortion-impact"-based solution and rate-distortion optimization assuming constant channel conditions by, on average, around 4dB and 2.5dB in "Foreman", 2dB and 1.5dB in "Coastguard", and 3.5dB and 1.5dB in "Mobile" in terms of PSNR. The improvement comes from the fact that our proposed solution schedules the packets and adapts the transmission strategies (i.e. adapting the power allocation) based on the heterogeneous characteristics of the multimedia packets as well as the time-varying channel conditions. We also notice that the rate-distortion optimization with constant channel conditions obtains higher received video quality than the “distortion-impact”-based solution. It shows that the characteristics (dependencies, distortion impacts and delay constants) of media packets play a very important role in improving the media quality. We further notice that the improvement of our proposed solution becomes much larger when the available resource becomes adequate. This is because, when the resource is scarce, all solutions only schedule the most important data, which has the highest distortion impact.

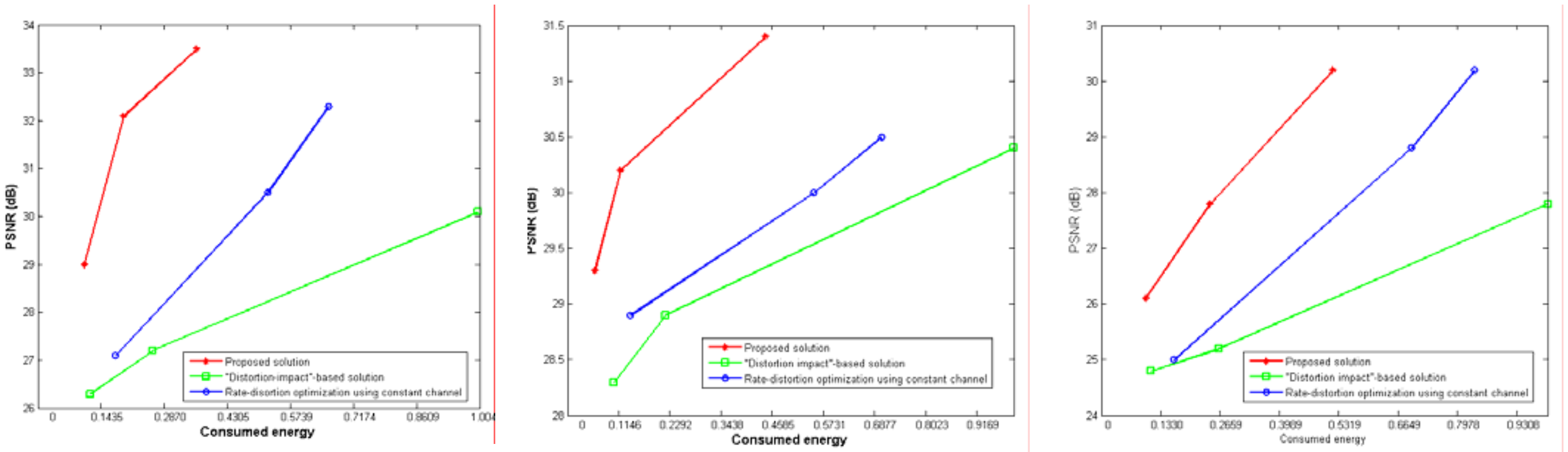

(a) “Foreman” (b) “Coastguard” (c) “Mobile”

Figure 8. PSNR-energy curve of “Foreman”, “Coastguard” and “Mobile” sequences for different transmission solutions

## B. Performance of cross-layer optimization with various delay constraints and GOP structures

In this section, we further compare the performance of the cross-layer optimization solutions for streaming the “Coastguard” video sequence. The cross-layer actions and wireless channel settings are the same as in Section A. However, we consider that the video sequence can be encoded using different GOP structures: 8 frames per GOP and 16 frames per GOP and tolerant different delay: a delay of 133ms (corresponding to half of the GOP duration) or a delay of 266ms (corresponding to the duration of one GOP). We compare our solution with the rate-distortion optimization using constant channels with different combinations of delay deadline and GOP structures. The PSNR- (normalized) energy curves are given in Figure 9. From this figure, we note that our proposed algorithm outperforms the rate-distortion optimization with constant channels under different GOP structures and delay deadlines, which confirms again the observation in Section A. We further notice that, by increasing the delay from 133ms to 266ms, the cross-layer optimization can improve, on average, 0.5dB in terms of PSNR. By increasing the number of frames in one GOP from 8 to 16, our solution can further improve 0.7dB in terms of PSNR. The improvement comes from the fact that, by increasing the delay, each media packet has more transmission opportunities and will be scheduled for transmission at a better channel condition. By increasing the number of frames in one GOP, the video sequence can be encoded more efficiently and there are fewer packets to be transmitted, which accordingly improves the video quality.

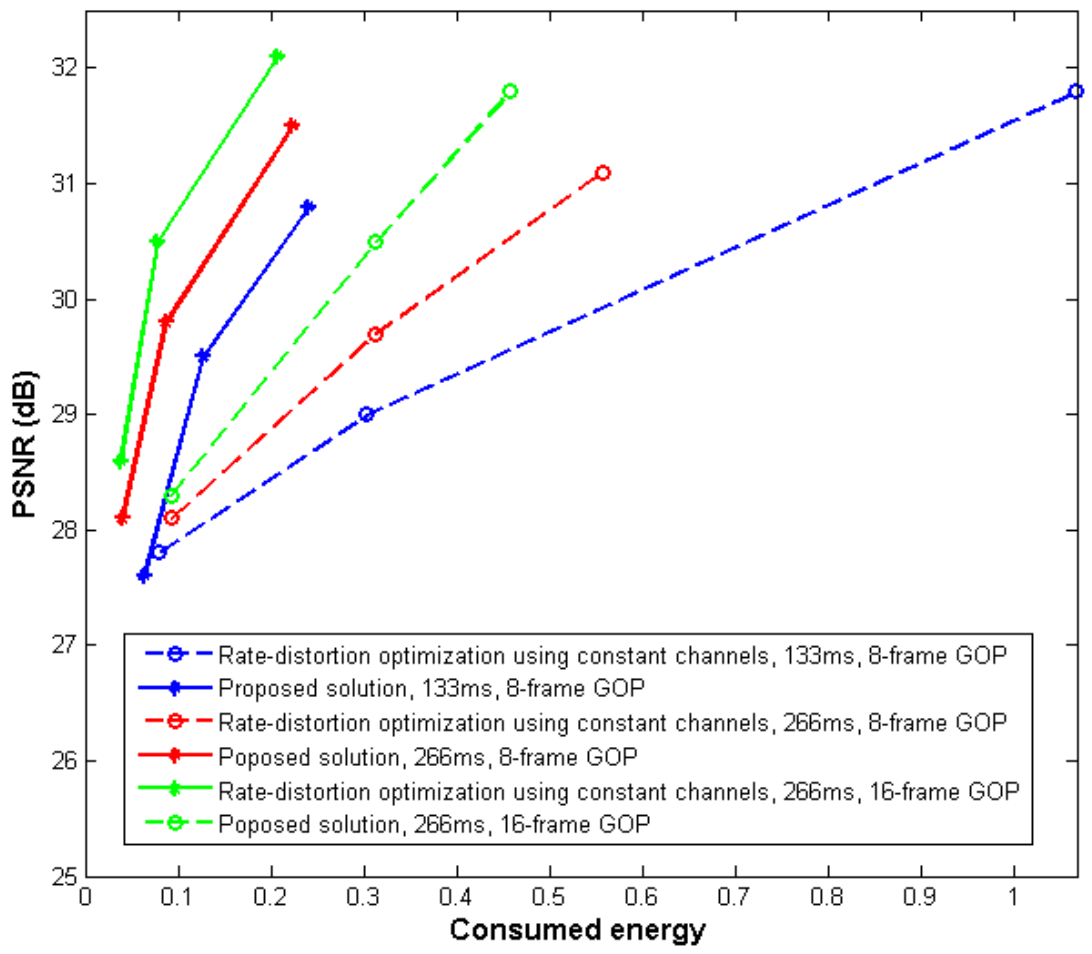

Figure 9. Video quality of “Coastguard” sequence with various delay deadlines and GOP structures

## C. Performance of cross-layer optimization under various packet loss characteristics

In this section, we evaluate the performance of our proposed cross-layer solution under different wireless channel conditions, exhibiting different packet loss patterns. In this experiment, we assume that the wireless user transmits the “Costguard” video sequence in CIF resolution 30Hz. The cross-layer action includes the packet scheduling at the APP layer and retransmission at the MAC layer which are adapted based on the experienced wireless channel conditions. We use the fixed power allocation and modulation and channel coding scheme at the PHY. The channel condition is represented by the Signal-to-Noise Ratio (SNR) varying from 10 dB to 30 dB and modelled as a FSMC [16] with 5 states. The transmission cost is the amount of time used for transmitting the video packets, as in Example 1. The cross-layer action is computed using the proposed “travelling state tree” algorithm and then performed under the wireless channels with two packet loss patterns: one has the average packet loss rate[13] 5% and the other one has average packet loss rate 10%.

Figure 10 shows the obtained video quality in terms of PSNR under these considered wireless channel conditions as well as the one without packet loss. We further compare our solution with the solution performing the rate-distortion optimization assuming the constant channel conditions as in RaDiO [7]. From this figure, we note that, when the wireless channel has 5% (10%) packet loss rate, the performance is degraded, on average, by 0.3dB (0.4dB) [14], compared to the idea wireless channel (without packet loss). This indicates that our proposed cross-layer solution is robust when the packet loss is small (e.g. <10%). Compared to the rate-distortion optimization solution with constant channel conditions, our proposed solution can still gain, on average, 0.7 dB (0.6 dB) in terms of PSNR when the wireless channel has 5% (10%) packet loss rate. In order to further improve the video quality, we need to explicitly take into account the packet loss for the cross-layer optimization and formulate it as a partially-observed MDP problem, which is part of our future research.

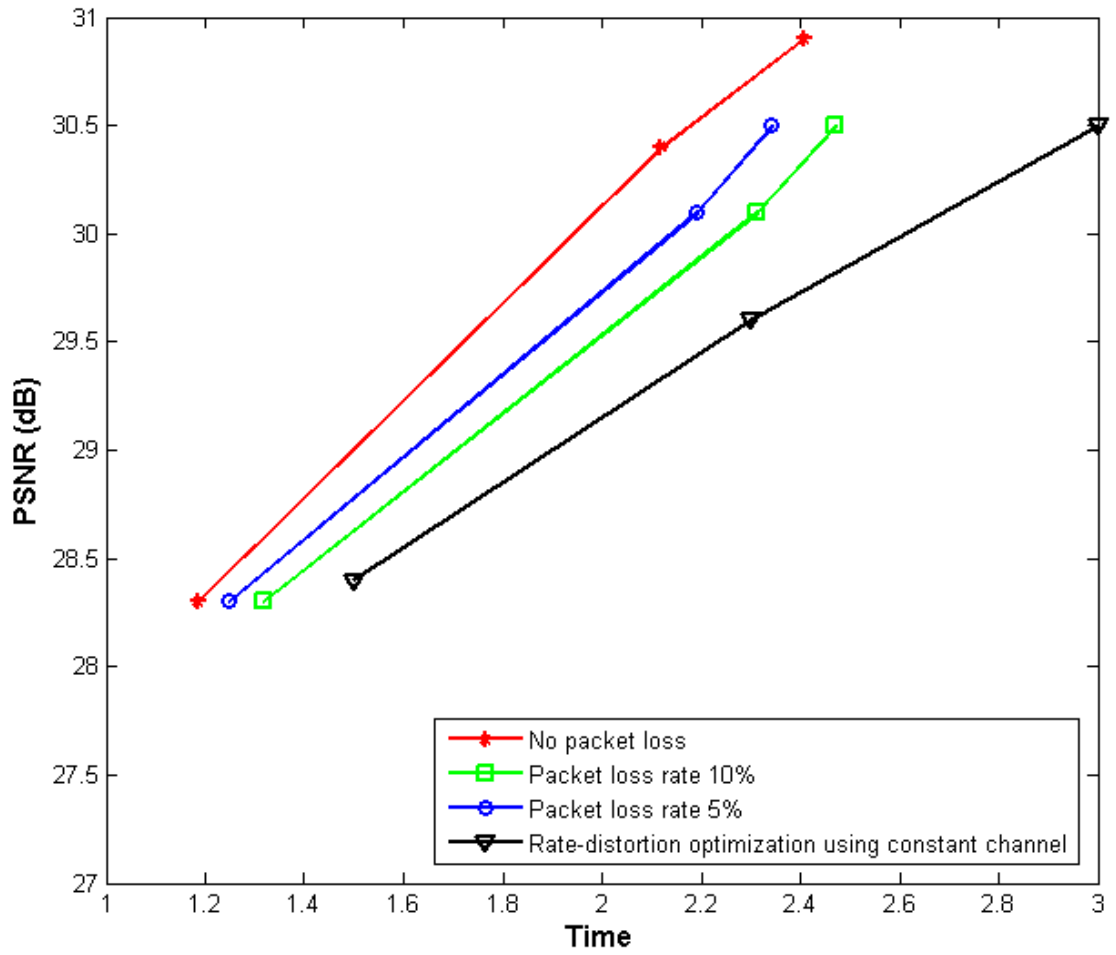

Figure 10. Video quality of "Coastguard" sequence under various packet loss rate

## D. Complexity evaluation

In this section, we compare the complexity of the proposed cross-layer optimization with the standard dynamic programming solution shown in Eqs. (5) and (6). We perform these two solutions for both independently decodable packets and interdependent packets. We randomly generate 20 independently decodable packets with

[13] The average packet loss is the packet loss computed when the channel state is steady.
[14] The degradation of less than 0.5dB in PSNR is often unnoticeable.

18 time slots and take the data from the "Coastguard" sequences as the interdependent packets. The complexities of both storage (in terms of the number of post-states whose value functions to be stored) and computation (in terms of the number of comparisons between packets in order to find optimal cross-layer actions) are listed in Table 2. From this result, we notice that our proposed solution significantly reduces the complexities by an order of 10 for independently decodable packets and by an order of $10^7$ for the interdependent packets. The complexity reduction is from the fact that our solution carefully prioritizes the packets according to the media characteristics while the standard dynamic programming does not consider this.

Table 2. Storage and computation complexities of cross-layer optimization

| | Standard dynamic programming | | Proposed solution | |
|---|---|---|---|---|
| | Number of post-states | Number of comparisons | Number of post-states | Number of comparisons |
| Independently decodable packets | 329 | 3258 | 26 | 385 |
| Interdependent packets ("Coastguard" video sequence) | $4.4\times10^{11}$ | $4.4\times10^{11}$ | 304 | 19381 |

## VII. CONCLUSIONS

In this paper, we formulate the problem of cross-layer optimization for delay-sensitive packetized media applications as a finite-horizon Markov decision process. Based on the heterogeneous characteristics of the media packets, we express the transmission priorities between packets as a DAG. Using the DAG expression, we are able to derive an optimal cross-layer solution as "*travelling the state tree*" by simply and recursively selecting the packet from the root of the priority graph having the highest marginal utility. Furthermore, from the DAG expression, we show that the complexity of the cross-layer optimization linearly increases with the disconnection degree of the priority graph but exponentially increases with the number of dependency states, both of which are determined by the characteristics of the media data. The simulation results show that the proposed cross-layer optimization solution significantly outperforms the start-of-art solutions which (partially) ignore the media characteristics and time-varying network conditions. Our solution also provides an upper bound of the performance obtained by the cross-layer optimization with delayed feedback such as RaDiO.

## APPENDIX

1. Proof of Theorem 1

We use backward induction to prove the statements in Theorem 1.

At time slot $d_j$, we have $U^{d_j}\left(s_j^{d_j}\right)=0, \forall s_j^{d_j}$.

In general, at time slot $t$, it is clear that $\pi_j^t\left(\left(0, h^t\right)\right)=0$ and $U_j^t\left(\left(0, h^t\right)\right)=0$, $\forall h^t \in \mathcal{H}$. For the state $s^t=\left(1, h^t\right)$,

$$
\begin{aligned}
\pi_j^t\left(\left(1, h^t\right)\right) & =\arg \max _{\pi_j^t}\left\{\left(q_j-\lambda \rho^t\left(l, h^t, a_j^{t,*}\right)\right) \pi_j^t+\alpha\left(1-\pi_j^t\right) \sum_{h^{t+1}} p\left(h^{t+1} \mid h^t\right) U^{t+1}\left(\left(1, h^{t+1}\right)\right)\right\} \\
& =\arg \max _{\pi_j^t}\left\{\left(q_j-\lambda \rho^t\left(l, h^t, a_j^{t,*}\right)\right) \pi_j^t+\left(1-\pi_j^t\right) \bar{u}_j^t\left(h^t\right)\right\} \\
& =\begin{cases}1 & \text{if } \left(q_j-\lambda \rho^t\left(l, h^t, a_j^{t,*}\right)\right)>\bar{u}_j^t\left(h^t\right) \\ 0 & o.w.\end{cases}
\end{aligned}
$$

where $a_j^{t,*}$ is the solution to $r\left(h^t, a_j^t\right)=l$. And

$$
U_j^t\left(\left(1, h^t\right)\right)=\max \left\{\left(q_j-\lambda \rho^t\left(l, h^t, a_j^{t,*}\right)\right), \bar{u}_j^t\left(h^t\right)\right\}.
$$

To prove (iii), we note that $\bar{u}^{d_j-1}\left(\alpha, h^{d_j-1}\right)=0$.

Then we have

$$
\begin{aligned}
& \bar{u}_j^{d_j-2}\left(\alpha, h^{d_j-2}\right)=\alpha \sum_{h^{d_j-1} \in \mathcal{H}} p\left(h^{d_j-1} \mid h^{d_j-2}\right) U_j^{d_j-1}\left(\left(1, h^{d_j-1}\right)\right) \\
& =\alpha E_{h^{d_j-1} \mid h^{d_j-2}} \max \left\{\left(q_j-\lambda \rho^{d_j-1}\left(l, h^{d_j-1}, a_j^{d_j-1,*}\right)\right), \bar{u}_j^{d_j-1}\left(\alpha, h^{d_j-1}\right)\right\} \\
& \geq \bar{u}_j^{d_j-1}\left(\alpha, h^{d_j-2}\right)=0
\end{aligned}
$$

where $E_{h^{d_j-1} \mid h^{d_j-2}} f\left(h^{d_j-1}\right)=\sum_{h^{d_j-1} \in \mathcal{H}} p\left(h^{d_j-1} \mid h^{d_j-2}\right) f\left(h^{d_j-1}\right)$.

Using the preceding inequality, we obtain for all $h^{d_j-3} \in \mathcal{H}$

$$
\begin{aligned}
& \bar{u}_j^{d_j-3}\left(\alpha, h^{d_j-3}\right)=\alpha E_{h^{d_j-2} \mid h^{d_j-3}} \max \left\{\left(q_j-\lambda \rho^{d_j-2}\left(l, h^{d_j-2}, a_j^{d_j-2,*}\right)\right), \bar{u}_j^{d_j-2}\left(\alpha, h^{d_j-2}\right)\right\} \\
& \geq \alpha E_{h^{d_j-2} \mid h^{d_j-3}} \max \left\{\left(q_j-\lambda \rho^{d_j-2}\left(l, h^{d_j-2}, a_j^{d_j-2,*}\right)\right), \bar{u}_j^{d_j-1}\left(\alpha, h^{d_j-2}\right)\right\} \\
& =\bar{u}_j^{d_j-2}\left(\alpha, h^{d_j-3}\right)
\end{aligned}
$$

Continuing in the same manner, we see that

$$
\bar{u}_j^t\left(\alpha, h^t\right) \geq \bar{u}_j^{t+1}\left(\alpha, h^t\right) \geq 0, \forall t, h^t .
$$

To prove (iv), we assume that $0 \leq \alpha_2 \leq \alpha_1 \leq 1$ and note that $\bar{u}_j^{d_j-1}\left(\alpha_1, h^{d_j-1}\right)=\bar{u}_j^{d_j-1}\left(\alpha_2, h^{d_j-1}\right)=0$.

$$\begin{aligned}&\overline{u}_j^{d_j-2}\left(\alpha_1,h^{d_j-2}\right)=\alpha_1 E_{h^{d_j-1}|h^{d_j-2}}\max\left\{\left(q_j-\lambda\rho_j^{d_j-1}\left(l_j,h^{d_j-1},a_j^{d_j-1,*}\right)\right),\overline{u}_j^{d_j-1}\left(\alpha_1,h^{d_j-1}\right)\right\}\\&\geq\alpha_2 E_{h^{d_j-1}|h^{d_j-2}}\max\left\{\left(q_j-\lambda\rho^{d_j-1}\left(l_j,h^{d_j-1},a_j^{d_j-1,*}\right)\right),\overline{u}_j^{d_j-1}\left(\alpha_2,h^{d_j-1}\right)\right\}\\&=\overline{u}_j^{d_j-2}\left(\alpha_2,h^{d_j-2}\right)\end{aligned}$$

By induction, at any time slot $t$, we have

$$\begin{aligned}&\overline{u}_j^{t}\left(\alpha_1,h^{t}\right)=\alpha_1 E_{h^{t+1}|h^{t}}\max\left\{\left(q_j-\lambda\rho_j^{t+1}\left(l_j,h^{t+1},a_j^{t+1,*}\right)\right),\overline{u}_j^{t+1}\left(\alpha_1,h^{t+1}\right)\right\}\\&\geq\alpha_1 E_{h^{t+1}|h^{t}}\max\left\{\left(q_j-\lambda\rho_j^{t+1}\left(l_j,h^{t+1},a_j^{t+1,*}\right)\right),\overline{u}_j^{t+1}\left(\alpha_2,h^{t+1}\right)\right\}\\&\geq\alpha_2 E_{h^{t+1}|h^{t}}\max\left\{\left(q_j-\lambda\rho_j^{t+1}\left(l_j,h^{t+1},a_j^{t+1,*}\right)\right),\overline{u}_j^{t+1}\left(\alpha_2,h^{t+1}\right)\right\}\\&=\overline{u}_j^{t}\left(\alpha_2,h^{t}\right)\end{aligned}.$$

■

2. Proof of Theorem 2

To prove the decomposition, we only need to prove that the dynamic programming in Eqs. (5) and (6) can be decomposed into $N$ dynamic programming solutions each of which corresponds to one single-packet cross-layer optimization.

$$U^{d_{\max}}\left(s^{d_{\max}}\right)=0=\sum_{j:t_j\leq d_{\max}\leq d_j}U_j^{d_{\max}}\left(s_j^{d_{\max}}\right)=\sum_{j:d_{\max}=d_j}U_j^{d_{\max}}\left(s_j^{d_{\max}}\right),\forall s^{d_j+1}.$$

$$U^t\left(s^t\right)=\max_{\pi^t,a^t}\left\{\sum_{j:t_j\leq t\leq d_j}q_j b_j^t\pi_j^t-\lambda\rho^t\left(\sum_{j,t_j\leq t\leq d_j}l_j b_j^t\pi_j^t,h^t,a^t\right)+\alpha\sum_{s^t}\prod_{j,t_j\leq t\leq d_j}p\left(b_j^{t+1}\mid b_j^t,\pi_j^t\right)p\left(h'\mid h\right)U^{t+1}\left(s^{t+1}\right)\right\}$$

Considering the linear transmission cost as shown in Eq. (8) and the decomposition of $U^{t+1}\left(s^{t+1}\right)=\sum_{j:t_j\leq t+1\leq d_j}U_j^{t+1}\left(s_j^{t+1}\right)$, we have

$$U^t\left(s^t\right)=\max_{\pi^t,a^t}\left\{\sum_{j:t_j\leq t\leq d_j}b_j^t\pi_j^t\left(q_j-\lambda\rho_j^t\left(l_j,h^t,a_j^t\right)\right)+\alpha\sum_{h'\in\mathcal{H}}\sum_{b_j^t,j,t_j\leq t\leq d_j}\prod_{j,t_j\leq t\leq d_j-1}p\left(b_j^{t+1}\mid b_j^t,\pi_j^t\right)p\left(h'\mid h\right)\cdot\sum_{j:t_j\leq t+1\leq d_j}U_j^{t+1}\left(s_j^{t+1}\right)\right\}.$$

By rearranging the above equation, we obtain

$$\begin{aligned}&U^t\left(s^t\right)=\max_{\pi^t,a^t}\left\{\sum_{j:t_j\leq t\leq d_j-1}\left\{b_j^t\pi_j^t\left(q_j-\lambda\rho_j^t\left(l_j,h^t,a_j^t\right)\right)+\alpha\sum_{h'\in\mathcal{H}}\sum_{b_j^t}p\left(b_j^{t+1}\mid b_j^t,\pi_j^t\right)p\left(h'\mid h\right)U_j^{t+1}\left(s_j^{t+1}\right)\right\}+\sum_{j:t=d_j}U_j^{d_j}\left(s_j^{d_j}\right)\right\}\\&=\sum_{j:t_j\leq t\leq d_j}U_j^t\left(s_j^t\right)\end{aligned}$$

The scheduling policy and transmission strategy are given by $\pi^{t,*}=\left\{\pi_j^{t,*}\right\}_{j:t_j\leq t\leq d_j}, a^{t,*}=\left\{a_j^{t,*}\right\}_{j:t_j\leq t\leq d_j}$. ■

3. Proof of Lemma 3

We first consider the case that $q_j \ge q_k$ and $d_j = d_k = d$. We need to prove that $\Delta u_j^t((1,h^t)) \ge \Delta u_k^t((1,h^t)), \forall h^t$. In the following, we use the matrix form. We define $\boldsymbol{y}(x) = x\mathbf{1} - \lambda\boldsymbol{\rho}$ where $\boldsymbol{\rho} = [\rho(h)]_{h\in\mathcal{H}}$.

At time slot $d$, we note that $\bar{\boldsymbol{u}}_j^d = \bar{\boldsymbol{u}}_k^d = 0$. Hence, $\Delta\boldsymbol{u}_j^d = \{\boldsymbol{y}(q_j) - \bar{\boldsymbol{u}}_j^d\}^+ = \{\boldsymbol{y}(q_j)\}^+$, and $\Delta\boldsymbol{u}_k^d = \{\boldsymbol{y}(q_k) - \bar{\boldsymbol{u}}_k^d\}^+ = \{\boldsymbol{y}(q_k)\}^+$ where $\{y\}^+ = \max\{y,0\}$. We note that

$$\begin{aligned}&\max\left\{\left(q_j - \lambda\rho_j^{t+1}\left(l_j,h^{t+1},a_j^{t+1,*}\right)\right),\bar{u}^{t+1}\left(\alpha,h^{t+1}\right)\right\}\\&= \bar{u}^{t+1}\left(\alpha,h^{t+1}\right) + \Delta u_j^{t+1}\left((1,h^{t+1})\right)\end{aligned}.$$

Then, we have

$$\bar{\boldsymbol{u}}_j^t = \alpha P\left\{\bar{\boldsymbol{u}}_j^{t+1} + \Delta\boldsymbol{u}_j^{t+1}\right\}.$$

where $P$ is the transition probability matrix of channel state.

Hence,

$$\bar{\boldsymbol{u}}^t(x) = \alpha P\bar{\boldsymbol{u}}^{t+1}(x) + \alpha P\Delta\boldsymbol{u}^{t+1}(x)$$

And

$$\Delta\boldsymbol{u}^t(x) = \left\{x\mathbf{1} - \lambda\boldsymbol{\rho} - \bar{\boldsymbol{u}}^t(x)\right\}^+.$$

In order to prove that $\Delta\boldsymbol{u}^t(q_j) \ge \Delta\boldsymbol{u}^t(q_k)$ if $q_j \ge q_k, d_j = d_k$, we only need to prove that $\Delta\bar{\boldsymbol{u}}^t(x)$ is an increasing function of $x$.

It is easy to show that

$$\mathbf{0} \le \frac{\partial\Delta\boldsymbol{u}^d(x)}{\partial x} = \frac{\partial\{\boldsymbol{y}(x)\}^+}{\partial x} \le \mathbf{1}$$

And

$$\mathbf{0} \le \frac{\partial\bar{\boldsymbol{u}}^{d-1}(x)}{\partial x} = \alpha P\frac{\partial\Delta\boldsymbol{u}^d(x)}{\partial x} \le \alpha\mathbf{1}.$$

In general, at time slot $t$,

$$\begin{aligned}&\frac{\partial\bar{\boldsymbol{u}}^t(x)}{\partial\boldsymbol{x}} = \alpha P\frac{\partial\bar{\boldsymbol{u}}^{t+1}(x)}{\partial\boldsymbol{x}} + \alpha P\frac{\partial\Delta\boldsymbol{u}^{t+1}(x)}{\partial\boldsymbol{x}}\\&= \alpha P\frac{\partial\bar{\boldsymbol{u}}^{t+1}(x)}{\partial\boldsymbol{x}} + \alpha P\frac{\partial\left\{x\mathbf{1} - \lambda\boldsymbol{\rho} - \bar{\boldsymbol{u}}^{t+1}(x)\right\}^+}{\partial\boldsymbol{x}}\end{aligned}$$

If $x\mathbf{1} - \lambda\boldsymbol{\rho} - \bar{\boldsymbol{u}}^{t+1}(x) > 0$, then

$$0 \le \frac{\partial \bar{\boldsymbol{u}}^t(x)}{\partial \boldsymbol{x}} \le \alpha \mathbf{1} ;$$

If $x\mathbf{1} - \lambda \boldsymbol{\rho} - \bar{\boldsymbol{u}}^{t+1}(x) \le 0$, then

$$0 \le \frac{\partial \bar{\boldsymbol{u}}^t(x)}{\partial \boldsymbol{x}} \le \alpha^2 \mathbf{1} ;$$

In summary, we have

$$0 \le \frac{\partial \bar{\boldsymbol{u}}^t(x)}{\partial \boldsymbol{x}} \le \alpha \mathbf{1} ;$$

It is easy to show that

$$\frac{\partial \Delta \boldsymbol{u}^t(x)}{\partial x} = \frac{\partial \left\{ x\mathbf{1} - \lambda \boldsymbol{\rho} - \bar{\boldsymbol{u}}^t(x) \right\}^+}{\partial x} \ge 0 ;$$

In other words, $\Delta \boldsymbol{u}^t(x)$ is an increasing function of $x$.

When $d_j < d_k$, we define $z = d - t$, then $\bar{\boldsymbol{u}}^z(x) = \bar{\boldsymbol{u}}^t(x)$ is an increasing function of $z$ and $\Delta \boldsymbol{u}^z(x)$ is a decreasing function as $z$ increases. That is,

$$\Delta \boldsymbol{u}^z(x) \ge \Delta \boldsymbol{u}^{z+d_k-d_j}(x) ;$$

Hence, $\Delta \boldsymbol{u}_j^t(q_j) \ge \Delta \boldsymbol{u}_j^{t-(d_k-d_j)}(q_j) \ge \Delta \boldsymbol{u}_k^{t-(d_k-d_j)}(q_k)$. ■

4. Proof of Theorem 3

We know that the optimal cross-layer actions can be found directly from the standard dynamic programming in Eqs. (5) and (6)., which is rewritten as follows:

$$U^t(s^t) = \max_{\pi^t, a^t} \left\{ \sum_{j:t_j \le t \le d_j} q_j b_j^t \pi_j^t - \lambda \rho^t \left( \sum_{j,t_j \le t \le d_j} l_j b_j^t \pi_j^t, h^t, a^t \right) + \alpha \sum_{s^t} \prod\nolimits_{j,t_j \le t \le d_j} p\left(b_j^{t+1} \mid b_j^t, \pi_j^t\right) p\left(h' \mid h\right) U^{t+1}\left(s^{t+1}\right) \right\}$$

Based on the computation of the post-state value function, the above dynamic programming becomes

$$U^t(s^t) = \max_{\pi^t, a^t} \left\{ \sum_{j:t_j \le t \le d_j} q_j b_j^t \pi_j^t - \lambda \rho^t \left( \sum_{j,t_j \le t \le d_j} l_j b_j^t \pi_j^t, h^t, a^t \right) + u^t\left(\tilde{B}^t(\pi^t), h^t\right) \right\}$$

where $\tilde{B}^t(\pi^t) = \{(1-\pi_j^t)b_j^t\}_{j,t_j \le t \le d_j}$ represents the remaining packets that have not been transmitted. Let us consider two packets $j'$ and $j''$. By symmetry, we only need to consider the following two scenarios: $j' \lhd j''$ and $j' \ntriangleleft j''$ where $j \ntriangleleft k$ means that packets $j'$ and $j''$ cannot be prioritized.

Case 1: $j' \lhd j''$

When $j' \lhd j''$, it can be easily shown from the definition of the transmission priority that

$$\begin{aligned} & q_{j'} + \lambda\left(\rho^t(l+x,h^t,a^t) - \rho^t(x,h^t,a^t)\right) + u^t\left(\left(0,1,\tilde{B}^t_{-j'-j''}\right),h^t\right) \\ & \ge q_{j''} + \lambda\left(\rho^t(l+x,h^t,a^t) - \rho^t(x,h^t,a^t)\right) + u^t\left(\left(1,0,\tilde{B}^t_{-j'-j''}\right),h^t\right) \end{aligned}$$

Where $\left(1,0,\tilde{B}^t_{-j'-j''}\right)$ is the post-state with $b^t_{j'} = 1$ and $b^t_{j''} = 0$ and $\tilde{B}^t_{-j'-j''}$ being the post-state of other packets.

By the transitivity of the transmission priority, we can conclude that we only need to compare the transmission order of the packets that cannot prioritized which is shown in case 2.

Case 2: $j' \ntriangleleft j''$

When $j' \ntriangleleft j''$, we cannot directly know which one should be transmitted first. However, since the dynamic programming is aimed to maximize the total utility, the packet with the highest marginal utility will be selected, i.e.

$$\begin{aligned} j^* &= \arg\max \begin{Bmatrix} q_{j'} + \lambda\left(\rho^t(l+x,h^t,a^t) - \rho^t(x,h^t,a^t)\right) + u^t\left(\left(0,1,\tilde{B}^t_{-j'-j''}\right),h^t\right), \\ q_{j''} + \lambda\left(\rho^t(l+x,h^t,a^t) - \rho^t(x,h^t,a^t)\right) + u^t\left(\left(1,0,\tilde{B}^t_{-j'-j''}\right),h^t\right) \end{Bmatrix} \\ &= \arg\max \begin{Bmatrix} q_{j'} + \lambda\left(\rho^t(l+x,h^t,a^t) - \rho^t(x,h^t,a^t)\right) + u^t\left(\left(0,1,\tilde{B}^t_{-j'-j''}\right),h^t\right) - u^t\left((B^t),h^t\right), \\ q_{j''} + \lambda\left(\rho^t(l+x,h^t,a^t) - \rho^t(x,h^t,a^t)\right) + u^t\left(\left(1,0,\tilde{B}^t_{-j'-j''}\right),h^t\right) - u^t\left((B^t),h^t\right) \end{Bmatrix}, \\ &= \arg\max\left\{\Delta u^t_{j'}, \Delta u^t_{j''}\right\} \end{aligned}$$

which corresponds to Phase 1, i.e. selecting the root packet in the DAG with the highest marginal utility. However, if the marginal utility $\Delta u^t_{j^*} \le 0$, packet $j^*$ will not be transmitted, which corresponds to Phase 2.

According to the standard dynamic programming, the state-value function is updated as follows:

$$\begin{aligned} & U^t\left((B^t,h^t)\right) = \sum_{k=1}^{k_{\max}} q_{j_k} - \lambda\rho\left(k_{\max}l,h^t\right) + \overline{u}^t\left(\left(PG^{k_{\max}},h^t\right)\right) \\ & U^t\left((\mathbf{0},h^t)\right) = 0 \end{aligned} . \blacksquare$$

5. State tree construction

**Algorithm 1**: State tree construction induced by a priority graph $PG$

**Input**: $PG$
**Output**: State tree $T$
**Initialization**: $V_T = \varnothing, E_T = \varnothing$ ; $Q_T = \varnothing$ ;
*Step 1*: $V_T = V_T \cup \{PG\}$ ;
*Step 2*: For any $v \in root(PG)$
    $E_T = E_T \cup \{PG/\{v\}\}$ ;
    Enqueue the graph $PG/\{v\}$ into $Q_T$
    End
*Step 3*: If the queue $Q_T$ is empty, stop and return $T$ ;
    Else dequeue a graph $PG'$ from $Q_T$ ;
    $PG = PG'$ ;
    End
*Ste 4*: Go back to step 1;